\documentclass[times, twoside]{Preprint}

\usepackage{blindtext}
\usepackage{graphicx}
\usepackage{dblfloatfix}
\usepackage{color, colortbl}
\usepackage{url}
\usepackage{upgreek}
\usepackage[breakable]{tcolorbox}

\definecolor{lightgray}{rgb}{0.95, 0.95, 0.95}
\definecolor{salmon1}{rgb}{1.0, 0.63, 0.48}
\definecolor{salmon2}{rgb}{1.0, 0.87, 0.68}
\definecolor{salmon3}{rgb}{1.0, 0.94, 0.84}
\definecolor{gray1}{rgb}{0.7, 0.7, 0.7}
\definecolor{gray2}{rgb}{0.85, 0.85, 0.85}
\definecolor{blue1}{rgb}{0.69, 0.77, 0.87}
\definecolor{blue2}{rgb}{0.94, 0.97, 1.0}
\definecolor{blue3}{rgb}{0.88, 1.0, 1.0}

\leadauthor{M. Ballester} 

\begin{document}

\setlength{\parindent}{1cm}

\title{Review and Novel Formulae for Transmittance and Reflectance of Wedged Thin Films on absorbing Substrates}

\shorttitle{Transmittance and Reflectance Formulae}

% ADJUST FONT FOR AUTHORS AND AFFILIATIONS
\setlength{\affilsep}{7 pt}
\renewcommand\Authfont{\color{black}\normalfont\sffamily\bfseries\fontsize{9}{11}\selectfont}
\renewcommand\Affilfont{\color{black}\normalfont\sffamily\fontsize{6.5}{8}\selectfont}
\renewcommand\Authands{, and }

\author[1,*]{Manuel Ballester}
\author[2]{Emilio Marquez}
\author[3]{John Bass}
\author[4]{Christoph Würsch}
\author[3]{\\Florian Willomitzer}
\author[1, 4]{Aggelos K. Katsaggelos}

\affil[1]{Department of Computer Sciences, Northwestern University, Evanston, IL 60208, USA}
\affil[2]{Department of Condensed-Matter Physics, Faculty of Science, University of Cadiz, 11510 Puerto Real, Spain}
\affil[3]{Wyant College of Optical Sciences, University of Arizona, Tucson, AZ 85721, USA}
\affil[4]{ETH Hoenggerberg Zürich, Stefano-Franscini-Platz 5, 8093, Switzerland}
\affil[4]{Department of Electrical and Computer Engineering, Northwestern University, Evanston, IL 60208, USA}

\affil[**]{Correspondence: manuelballestermatito2021@u.northwestern.edu}

\maketitle

%TC:break Abstract
%the command above serves to have a word count for the abstract
\begin{abstract}

Historically, spectroscopic techniques have been essential for studying the optical properties of thin solid films. However, existing formulae for both normal transmission and reflection spectroscopy often rely on simplified theoretical assumptions, which may not accurately align with real-world conditions. For instance, it is common to assume (1) that the thin solid layers are deposited on completely transparent thick substrates and (2) that the film surface forms a specular plane with a relatively small wedge angle. While recent studies have addressed these assumptions separately, this work presents an integrated framework that eliminates both assumptions simultaneously. In addition, the current work presents a deep review of various formulae from the literature, each with their corresponding levels of complexity. Our review analysis highlights a critical trade-off between computational complexity and expression accuracy, where the newly developed formulae offer enhanced accuracy at the expense of increased computational time. Our user-friendly code, which includes several classical transmittance and reflectance formulae from the literature and our newly proposed expressions, is publicly available in both Python and Matlab at this \href{https://drive.google.com/drive/folders/1Mv0p9or5ePowgt37yitNnw2Xe449IFTG?usp=sharing}{link}.

\end {abstract}
%TC:break main
%the command above serves to have a word count for the abstract

\vspace{-3mm}

\section{Introduction}
\label{sec:intro}

Thin solid films are essential in a wide range of modern industries, especially in the development of efficient transistors and photodiodes, as well as in the fabrication of protective metal and dielectric coatings. Thin film technologies directly impact the performance and efficiency of several everyday devices, including active matrix LCDs, photovoltaic systems, flash memory chips, photonic integrated circuits, and semiconductor batteries \cite{marquez2021optical, birkett2016optical, rodriguez2024power, marquez2023mid, lopezenhanced}. It is important to remark that the specific conditions under which thin films are prepared have a significant impact on their final properties. Factors such as deposition techniques, preparation time, growth temperature, and working pressure all play critical roles in determining the characteristics of the resulting films \cite{marquez1992calculation, marquez2019influence, rodriguez2023impact}. Consequently, a precise analysis of their optoelectronic properties prior to mass production is crucial. 
\begin{figure*}[t!]
  \centering
   \includegraphics[width=\linewidth]{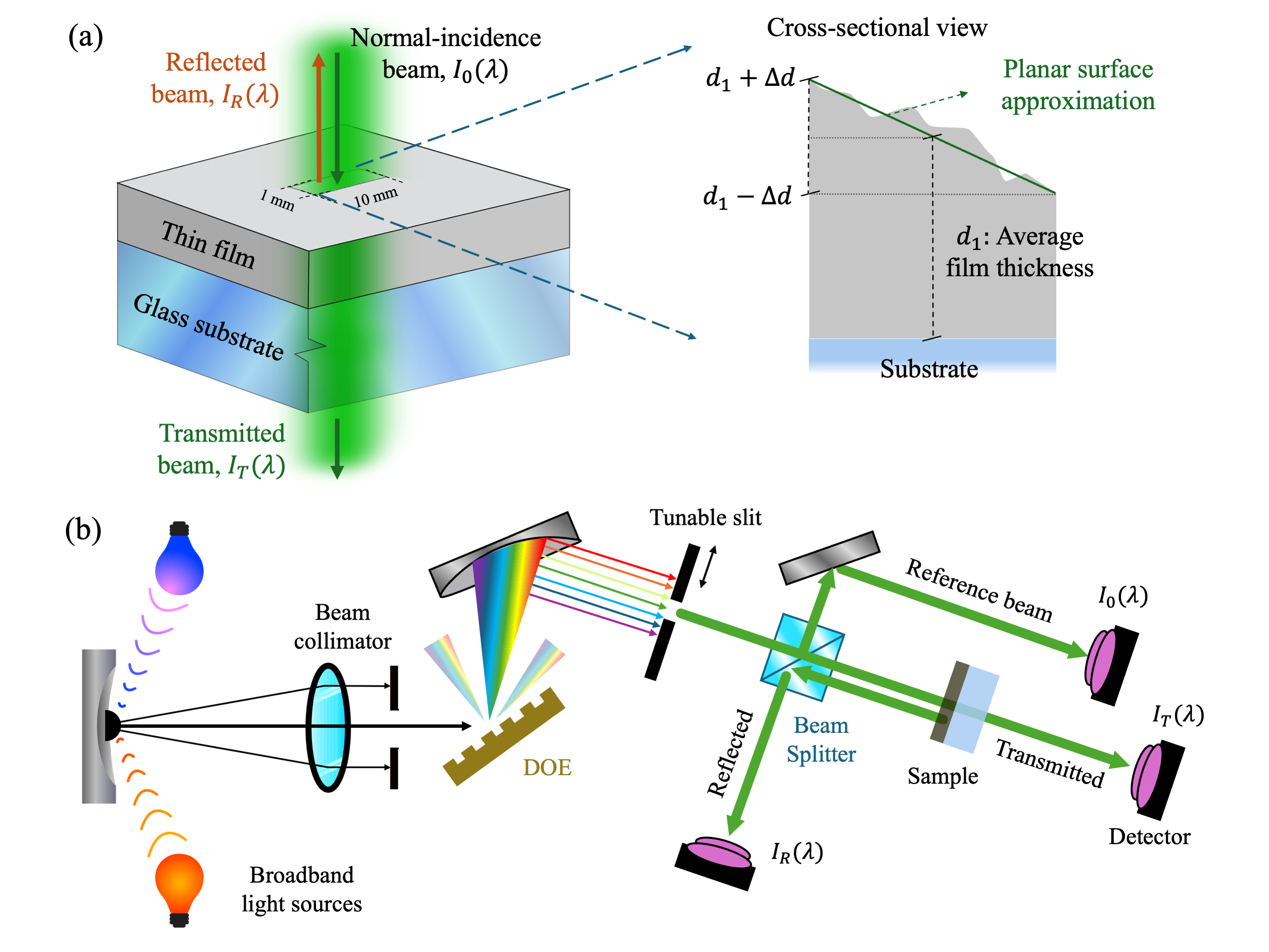}

   \caption{(a) Geometrical model of the sample under study: A thin solid film is deposited on top of a glass substrate. Inset with the cross-sectional view of the film in the spot illumination area. The thin film surface is approximated as a plane with a certain wedge parameter $\Delta d$. (b) Diagram of the transmission and reflection spectroscopy setup. Note the two broadband light sources (for UV and VIS-NIR), the diffraction optical element, the tunable slit that produces a quasi-monochromatic beam, and the light beam splitter.}
   \label{fig:experiment}
\end{figure*}

Thin films, with thicknesses often ranging from nanometers to several micrometers, are commonly deposited onto thick glass substrates (often on the millimeter scale), as seen in Fig. \ref{fig:experiment}a. Our main objective is to determine the optical properties of the film, specifically the refractive index $n_1(\lambda)$ and the extinction coefficient $\kappa_1(\lambda)$, as functions of the wavelength $\lambda$, within a broad spectral range of interest, typically in the UV-Vis-NIR region. While there are multiple methods for optically characterizing these thin solid films, \textit{spectroscopic ellipsometry} often emerges as the preferred choice due to its high accuracy. This technique measures how the polarization state of an incident beam changes upon reflection on the layer surface \cite{azzam1987ellipsometry,tompkins2015spectroscopic}. Spectroscopic ellipsometry techniques provide comprehensive data collection through different reflection angles and polarization states, which allows for highly precise thin-film material characterization. It should be noted that this technique is particularly effective for analyzing complicated sample structures, including multilayered \cite{bader1998transmission, petrik2006optical, jena2020evolutionary} or anisotropic materials \cite{den1971ellipsometry, fodor2017spectroscopic}.

Despite the significant benefits of ellipsometry techniques, it is still very common to perform the optical characterization of simple thin films using measurements of normal transmittance and/or reflectance \cite{marquez2023complex}. Several compelling factors contribute to the continued reliance on these latter measures. First, transmission and reflection spectrophotometers are indispensable instruments with applications in a wide range of fields, including biology and chemistry \cite{gore2000spectrophotometry}. Second, the widespread availability of these instruments in research institutes and industries makes them a practical choice. Third, these instruments are generally more economical than spectroscopic ellipsometers, offering a cost-effective alternative for many laboratories \cite{bosch1998spectrophotometry}. In addition, many commercial instruments are designed to integrate both optical transmittance and reflectance measurements, facilitating efficient data acquisition \cite{marceaumeasurement}. Fourth, collecting a single set of light intensity data at normal incidence is generally easier, simpler, and faster than collecting a comprehensive ellipsometric dataset. Fifth, ellipsometric data strongly depend on surface roughness \cite{aspnes1982optical, stroud1998effective, adachi2000optical, lehmann2014thin}, as optically rough surface layers lead to significant depolarization upon reflection. In contrast, the total transmitted and reflected intensity measurements are less influenced by this factor \cite{pradeep2010determination}. It should also be noted that, in spectral ranges where the films exhibit medium-to-low absorption, most of the incident light is transmitted, enabling highly sensitive and accurate optical characterizations based on these transmission measurements. In contrast, for characterizations within the strong absorption regions of a film material, reflection measurements (such as those performed in reflectometry or ellipsometry) offer more advantages. 

A representative schematic of a double-beam spectrophotometer can be found in Fig. \ref{fig:experiment}b. A lamp emits broad-spectrum light. Typically, a deuterium arc lamp is used for the UV region, and a tungsten-halogen lamp for the Vis-NIR spectrum. This light is then dispersed by a diffraction optical element (DOE), as seen in the figure. Consequently, it passes through a slit, allowing for a controllable spectral bandwidth that usually ranges from 0.1 to 10 nm, which determines the coherent length of the beam. Various apertures are available to adjust the beam diameter, generally in the millimeter range. The normal-incidence reflected and transmitted light intensities, denoted as $I_\text{T}$ and $I_\text{R}$, are then measured by a photodetector. Double-beam instrumentation splits the input beam into two parts: one to analyze the sample and the other to serve as a reference, measuring the baseline intensity $I_0$ with a separate sensor. Variations in the light source intensity are then recorded simultaneously in both the reference and sample beams, allowing an accurate intensity ratio for transmission $T=I_\text{T}/I_0$ and reflection $R=I_\text{R}/I_0$. Then, $N$ sequential transmittance and reflectance experimental measurements, denoted $\{T^\text{exp}(\lambda_i), R^\text{exp}(\lambda_i)\}$, are obtained at discrete specific wavelengths $\lambda_i$, $i \in \{1,2,...,N\}$, with a typical step size of a few nanometers. Our goal is then to develop a computational method that determines the two optical constants, $n_1(\lambda_i)$ and $\kappa_1(\lambda_i)$, from that experimental dataset.

Since the proliferation of advanced spectroscopic techniques, numerous analytical expressions have been formulated for the spectral transmission and reflection of thin-film samples, denoted $T^\text{theory}(\lambda; n_1, \kappa_1)$ and $R^\text{theory}(\lambda; n_1, \kappa_1)$. The formulae were then used to computationally fit the real-world optical measurements, finding the most reasonable values of $(n_1, \kappa_1)$ that match the experiment. However, these theoretical expressions generally relied on rough assumptions or approximations. The approximations introduced were necessary in order to simplify the analysis and reduce computational complexity, particularly given the limited computing resources available at the time. Among the simplifications that still persist nowadays, two are particularly critical: (1) the glass substrate is presumed to be completely transparent throughout the spectral range of interest, and (2) the film surface is regarded as a specular plane with a minor wedge angle, usually only a few nanometers in height.

Current instruments are capable of detecting very subtle variations in light intensity measurements, revealing the existing weak substrate absorption, because of the presence of inherent glass absorption at specific spectral ranges or impurities in the substrate. Additionally, although it is true that most films are conveniently prepared to have quasi-uniform thicknesses, more `exotic' thin-film samples may display complex surface geometries \cite{ohlidal2011measurement}. Such variations in thickness can lead to a pronounced wedge effect on the illuminated spot under analysis (see Fig. \ref{fig:experiment}a). It should be highlighted that the modern literature has already addressed simplifications (1) and (2) separately, as explained in the next section in further detail. In contrast, the current work presents a unified and comprehensive theoretical model that addresses both problems \textit{simultaneously}, bridging the gap in the current literature. Our code, written both in Python 3.0 \cite{python_3} and Matlab 2021a \cite{MATLAB_R2021a}, is available to the public at this \href{https://drive.google.com/drive/folders/1Mv0p9or5ePowgt37yitNnw2Xe449IFTG?usp=sharing}{link}. Please note that this code also includes a collection of previously established formulae in this field, accounting for the results of $T^\text{theory}$ and $R^\text{theory}$ under different approximations.

\section{Historical context and previous works}

The study of the physics of thin films has profoundly influenced the fields of optics and photonics, particularly in the analysis of optical interference effects during the nineteenth century \cite{hecht2012optics, born2013principles}. Relevant advancements in this field include the development of the Fabry-Pérot interferometer, the analysis of the etalon effect, and the formulation of Airy's equations. These contributions have facilitated advances in instruments for precise wavelength determination and enabled the development of accurate optical components. In addition, the examination of absorption bands in these thin layer materials significantly advanced atomic physics in the early twentieth century, improving our understanding of material properties at the quantum level. During the latter half of the twentieth century, there was a proliferation of accurate optoelectronic analyses of film semiconductor materials. The key advances that propelled these characterizations included \textit{(i)} the introduction of \textit{automated} spectrophotometers \cite{simoni2003classic, vedam1998spectroscopic}, \textit{(ii)} the integration of modern computer technologies and numerical methods, and \textit{(iii)} the continuous development and refinement of film preparation techniques \cite{butt2022thin}. 

%Additionally, it should be noted that many components in modern optical setups depend on thin-film technologies, such as anti-reflection coatings, thin-film polarizers, bandpass filters, dichroic beamsplitters, or multilayer mirrors. The role of photonic devices, including quantum optoelectronic devices, is paramount for efficient signal transmission and reception in our digital world. As we move toward a future with increased reliance on high-speed data transfer and complex communication systems, the importance of these technologies will only grow, underscoring the need for rigorous analysis and optimization of thin-film materials and structures.

\subsection{Computational methods to find optical properties}
It is essential to review the historical context of well-established computational methods that perform optical characterization. We mentioned above a straightforward approach that consists of directly fitting the experimental transmittance (or reflectance) measurements with the derived theoretical formulae. By minimizing the least-squares error between the experimental data and our theoretical models, we find the optimal values for the optical functions, $n_1$ and $\kappa_1$. This approach is commonly known as ``inverse synthesis'' or ``reverse engineering'', and involves solving a complicated global optimization problem \cite{dobrowolski1983determination, buffeteau1989thin, kamble2011determination, gao2012reverse, jafar2016retrieval,  tejada2019determination, alonso2024comprehensive}. 

In contrast, Hall and Ferguson (1955) \cite{hall1955optical}, Lyashenko and Miloslavskii (1964) \cite{Lyashenko}, and Manifacier \textit{et al.} (1975) \cite{manifacier1976simple} developed an alternative approach to find optical properties, initially known as the method of ``successive iterations''. This method does not operate for all the discrete measure wavelengths, but rather works for those particular wavelengths at which the minimum and maximum thin-film interferences occur. This algorithm requires the calculation, during the intermediate steps, of the lower and upper \textit{envelopes} of the spectral transmission and/or reflection curves. 

Ryno Swanepoel significantly refined this technique, introducing more precise formulae in two seminal works in 1983 \cite{swanepoel1983determination} and 1984 \cite{swanepoel1984determination}, respectively for uniform and wedged films. Subsequently, the algorithm was universally renamed as the \textit{Swanepoel method} to his honor. Initially limited to transmission measures, this methodology was then expanded to include reflection spectra by Minkov \textit{et al.} in a series of studies \cite{minkov1989calculation, minkov1989method, minkov1991computation, minkov1991errors, minkov1995comparative}. Since then, numerous significant works have further advanced this method for both transmission and reflection spectroscopy \cite{marquez1995optical, gonzalez1998derivation, ruiz1998new, richards2004determination, shaaban2012validity, dorranian2012optical, jin2017improvement, minkov2023increasing, ruiz2001method, filippov2000method}.

A comparison between the \textit{inverse synthesis} and \textit{envelope method} \cite{liu2011comparison, ballester2022comparison} indicates that, although inverse synthesis typically generally achieves greater precision, it comes with significantly higher computational costs. In addition to these two traditional methods, new research directions are emerging. For example, a novel deep learning (DL) technique has shown promise in accurately predicting optical properties of thin solid films from reflection \cite{wang2023measuring} and transmission \cite{ballester2024optical} spectra. Furthermore, a hybrid method has been proposed that combines the traditional Swanepoel method with DL enhancements \cite{ballester2023deep}. In that work, a neural network automatically determines the envelopes of a transmission spectra, and then an automatic algorithm determines the optical constants using the information contained in the envelopes. Another approach recently proposed by our group is a numerical method that compares the experimental transmittance with the transmittance generated by a \textit{numerical simulator} \cite{bass2024angular}. The simulator closely replicates wave propagation under the realistic conditions of a spectrophotometer, effectively acting as a digital twin of the actual setup.

\subsection{Existing transmission and reflection formula}

Revisiting now the early theoretical analysis of transmission and reflection formulae for thin solid films, it is essential to highlight the contributions of A.W. Crook in 1948 \cite{crook1948reflection} and F. Abelès in the 1950s \cite{abeles1950determination, abeles1953facteurs, abeles1957optical}. They independently laid the foundation for the transmission and reflection analysis in stratified media. The compilation work in the original Heavens book ``Optical Properties of Thin Solid Films'' from 1955 \cite{mullen1956optical, heavens1991optical} further consolidated the theoretical formulations employed in this field.

Despite these incipient accurate analyzes, it was common practice during that period to simplify the theoretical expressions and assume that the film was deposited on an \textit{infinitely thick} substrate \cite{manifacier1976simple, nestell1972derivation, perez1997nuevos}. Observe that the approximation overlooks the impact of back-reflection at the later substrate-air interface. The approximation significantly simplified the formulae to favor a straightforward optical characterization employing the envelope method \cite{manifacier1976simple}. In 1983 and 1984, the two aforementioned seminal works by Swanepoel \cite{swanepoel1983determination, swanepoel1984determination} formally challenged the notion of neglecting the actual thickness of the substrate. These studies demonstrated that such simplifications could result in errors in transmittance values of up to 3-4\%. Subsequent studies highlighted the importance of accounting for substrate back-reflection in reflectance analyses \cite{cesaria2014realistic}.

In 1971, Potapov and Rakov proposed a pioneer algorithm to account for the effect of \textit{slightly absorbing} thick substrates \cite{potapov1971determination} on the transmittance and reflectance of samples with uniform films. Another relevant subsequent work on the topic was later reported by Vriens and Rippens in 1983 \cite{vriens1983optical}. A later work by Swanepoel in 1989 \cite{swanepoel1989transmission} meticulously derived the comprehensive formulae describing the effect of a quasi-coherent light source. It should be noted that these close-form friendly expressions and methodology proposed by Swanepoel have been adopted in the present work. In 1997, Kotlikov and Terechenko \cite{kotlikov1997study, vargas2007closed} independently reached findings identical to those of Swanepoel, although in the context of antireflection coatings. Continuing the discussion on substrate absorption, it is important to highlight a study \cite{nichelatti2002complex} by Nichelatti (2002), which introduces a very efficient numerical technique to directly characterize substrate optical properties through spectroscopic measurements of that isolated substrate (explained in Section 9). In 2010, Barybin and Shapovalov \cite{barybin2010substrate} offered an alternative derivation of the transmission and reflection formulae for uniform films using the matrix formalism, specifically accounting for the effects of highly absorbing substrates.

The analysis of transmission and reflection formulae has continuously evolved up to the present day in different directions, accommodating increasingly complex sample characteristics. For example, some studies consider advanced thin film geometries \cite{kotlikov2009thickness, nevcas2009reflectance, franta2013advanced, baek2006determination} or deal with multiple thin-layer configurations \cite{prentice2000coherent, katsidis2002general}. In particular, a new significant contribution came from Ruiz-Perez (RP) \textit{et al.} \cite{ruiz2020optical} in 2020, adapting the transmission formula to account for films with a \textit{pronounced wedge} deposited on transparent substrates. The RP formula accounts for `exotic' films with a high-wedge condition (explained below in detail), which leads to significant changes in the shape of the spectrum, shrinking the characteristic oscillations of the transmission curves. This particular transmission expression has been successfully used in the recent literature and was named as the ``universal transmission formula'' \cite{ballester2022energy, ballester2022application}. Built upon the RP approach, our current work presents both transmission and reflection formulae, which are now able to account not only for complex films with a high wedge angle but also for thick absorbing glass substrates. Table \ref{tab:summary} summarizes the contributions of several relevant formulae from the literature, together with their associated approximations and their corresponding degree of precision.

\renewcommand{\arraystretch}{1.5}
\begin{table*}[!h]
\centering
\begin{tabular}{|c|c|c|c|c|c|}
\hline
\textbf{Formula} & \textbf{Equation} & \textbf{Work} & \textbf{Coherence} & \textbf{Uniform} & \textbf{Approximations} \\
\hline
$T$& Eq. \ref{eq:coherence_T} & Established \cite{yeh1990optical} & $\infty$& Yes& Exact\\
\hline
$R$& Eq. \ref{eq:coherence_R} & Established \cite{yeh1990optical} & $\infty$& Yes& Exact\\
\hline
$T_\ell$& Eq. \ref{eq:T_quasi} & Swanepoel 1989 \cite{swanepoel1989transmission} & $2 d < L<2 d_\text{s}$& Yes& Exact\\
\hline
$R_\ell$& Eq. \ref{eq:R_quasi} & Swanepoel 1989 \cite{swanepoel1989transmission} & $2 d < L<2 d_\text{s}$& Yes& Exact\\
\hline
$T^\text{new}_{\Delta d}$& Eq. \ref{eq:T_delta} & Current & $2 d < L<2 d_\text{s}$& No& Exact\\
\hline
$R^\text{new}_{\Delta d}$& Eq. \ref{eq:R_delta} & Current & $2 d < L<2 d_\text{s}$& No& Exact\\
\hline
$T^\text{RP20}_{\Delta d}$& Eq. \ref{eq:T_JJ} & Ruiz Perez 2020 \cite{ruiz2020optical} & $2 d < L<2 d_\text{s}$& No& $\kappa_2 = 0$ \\
\hline
$R^\text{Mink89}$& Eq. \ref{eq:R_Minkov89} & Minkov 1989 \cite{minkov1989calculation} & $2 d < L<2 d_\text{s}$& No& $\kappa_2 = 0$ \\
\hline
$T^\text{Swan83}$& Eq. \ref{eq:T_swan83} &  Swanepoel 1983 \cite{swanepoel1983determination} & $2 d < L<2 d_\text{s}$& Yes& $\kappa_2 = 0$ and $\kappa_1 \ll n_1$ \\
\hline
$T^\text{Swan84}_{\Delta d}$& Eq. \ref{eq:T_swan84} & Swanepoel 1984 \cite{swanepoel1984determination} & $2 d < L<2 d_\text{s}$& No& $\kappa_2 = 0$ and $\kappa_1 \ll n_1$ \\
\hline
$R^\text{RP01}$& Eq. \ref{eq:R_RP01_swan} & Ruiz Perez 2001 \cite{ruiz2001method} & $2 d < L<2 d_\text{s}$& Yes& $\kappa_2 = 0$ and $\kappa_1 \ll n_1$ \\
\hline
$R^\text{RP01}_{\Delta d}$& Eq. \ref{eq:R_RP01_swan_delta} & Ruiz Perez 2001 \cite{ruiz2001method} & $2 d < L<2 d_\text{s}$& No& $\kappa_2 = 0$ and $\kappa_1 \ll n_1$ \\
\hline
\hline
$T_\text{s}^\text{approx}$ & Eq. \ref{eq:R_RP01_swan_delta} & Established \cite{marquez2021optical} & $L<2 d_\text{s}$& Yes & $\kappa_2 = 0$ \\
\hline
$R_\text{s}^\text{approx}$ & Eq. \ref{eq:R_RP01_swan_delta} & Established \cite{gonzalez1998derivation} & $L<2 d_\text{s}$& Yes & $\kappa_2 = 0$ \\
\hline
$T_\text{s}$ & Eqs. \ref{eq:Ts}, \ref{eq:n2_from_TR}  & Nichelatti 2002 \cite{nichelatti2002complex} & $L<2 d_\text{s}$& Yes & Exact \\
\hline
$R_\text{s}$ & Eqs. \ref{eq:Rs}, \ref{eq:n2_from_TR} & Nichelatti 2002 \cite{nichelatti2002complex}& $L<2 d_\text{s}$& Yes & Exact \\
\hline
\end{tabular}
\caption{Set of transmittance and reflection expressions analyzed in the present work, each with their corresponding levels of accuracy and approximations.}
\label{tab:summary}
\end{table*}

\section{Theoretical background}
\label{secbackground}
The present study assumes that the thin film is made of a homogeneous material with isotropic properties. We describe the complex dielectric function (the electric permittivity) across the stratified structure as 
\begin{equation}
\boldsymbol{\epsilon}(z) = \left\{
\begin{array}{l l l}
\boldsymbol{\epsilon_{0}}, & z < 0 & \text{(Air)} \\
\boldsymbol{\epsilon_{1}}, & 0 < z < d_1 & \text{(Film)} \\
\boldsymbol{\epsilon_{2}}, & d_1 < z < d_1 + d_{\text{s}} & \text{(Substrate)} \\
\boldsymbol{\epsilon_{0}}, & d_1 + d_{\text{s}} < z & \text{(Air)}
\end{array}
\right.
\end{equation}
The bold notation represents the complex nature of the functions. In this formula, $d$ is the thickness of the film and $d_\text{s}$ the thickness of the substrate. The initial sections 2-4 assume a film of uniform thickness $\Delta d = 0$. Later on, we explore films with a certain tilt, including those with the high wedge condition $\Delta d > \lambda/(4 n)$, which requires a careful and distinct mathematical treatment, as discussed in detail in \cite{ruiz2001method}. Although multiple surface geometries could be considered for the film, a plane surface with a potential wedge parameter $\Delta d$ is reasonably assumed for the small illuminated area. This illuminated spot is often represented as a rectangular region with only a few millimeters of width and height . This planar assumption can be thought of as a Taylor first-order approximation of the actual much more intricate film surface (see Fig. \ref{fig:experiment}a).

Maxwell equations express a relationship between the electric and magnetic fields \cite{hecht2012optics} , denoted $\vec{E}$ and $\vec{H}$. Note that an external electric field $\vec{E}$ can influence an internal electric polarization $\vec{P}$ within a dielectric or semiconductor material, described as 
\begin{equation}
    \vec{P} = \epsilon_0 \boldsymbol{\chi}^{(1)} \vec{E} + \epsilon_0 \boldsymbol{\chi}^{(2)} \vec{E}^2 + \epsilon_0 \boldsymbol{\chi}^{(3)} \vec{E}^3 + ...
\label{eq:flux_density}
\end{equation} 
The electric flux density within the material is then given by $\vec{D} = \epsilon_0 \vec{E} + \vec{P}$. For non-magnetic materials, the magnetic flux density is simply $\vec{B} = \mu_0 \vec{H}$, where $\mu_0$ is the magnetic permeability in vacuum. In Eq. \ref{eq:flux_density}, the Taylor approximation around $\vec{E}=0$ was used to approximate an arbitrary polarization function, considering complex susceptibility coefficients $(\boldsymbol{\chi}^{(1)}, \boldsymbol{\chi}^{(2)}, \boldsymbol{\chi}^{(3)},...)$. In practice, most materials exhibit a linear response, although second- and third-order effects have also been studied in depth \cite{nonlinear} since the invention of laser in 1960, especially in the context of high intensity light beams. Our research focuses exclusively on linear materials, where $\vec{D} = \epsilon_0 (1 + \boldsymbol{\chi}^{(1)}) \vec{E} = \epsilon_0 \boldsymbol{\epsilon_r} \vec{E} = \boldsymbol{\epsilon} \vec{E}$. In these particular cases, the complex relative permittivity $\boldsymbol{\epsilon_r}$ fully describes the optoelectric properties of the material.

In linear non-magnetic isotropic film materials, the complex refractive index is defined as $\mathbf{n}(z) = \sqrt{\boldsymbol{\epsilon_r}} = n(z) + i \kappa(z)$. Here, $n$ denotes the real part of the refractive index, proportional to the phase velocity of light through the medium. The extinction (or attenuation) coefficient $\kappa$  is closely related to the absorption of the medium. The complex dielectric function reveals key information about the electronic transitions of the material. In turn, this allows us to deduce several critical aspects relevant to the film industry. For instance, it provides estimations of the band gap energy \cite{ballester2022energy}, the material conductivity \cite{stadler2010analyzing}, the dissipation factor \cite{wahab2012study}, the compound stoichiometry \cite{marquez2023optical}, and the level of structural disorder \cite{marquez2021optical}. Understanding structural disorder is particularly important for the analysis of the physical properties of amorphous materials and the identification of defects in crystalline structures.

\section{Light propagation through a sample}
\label{sec:coherent-uniform}

\begin{figure}[t!]
  \centering
   \includegraphics[width=\linewidth]{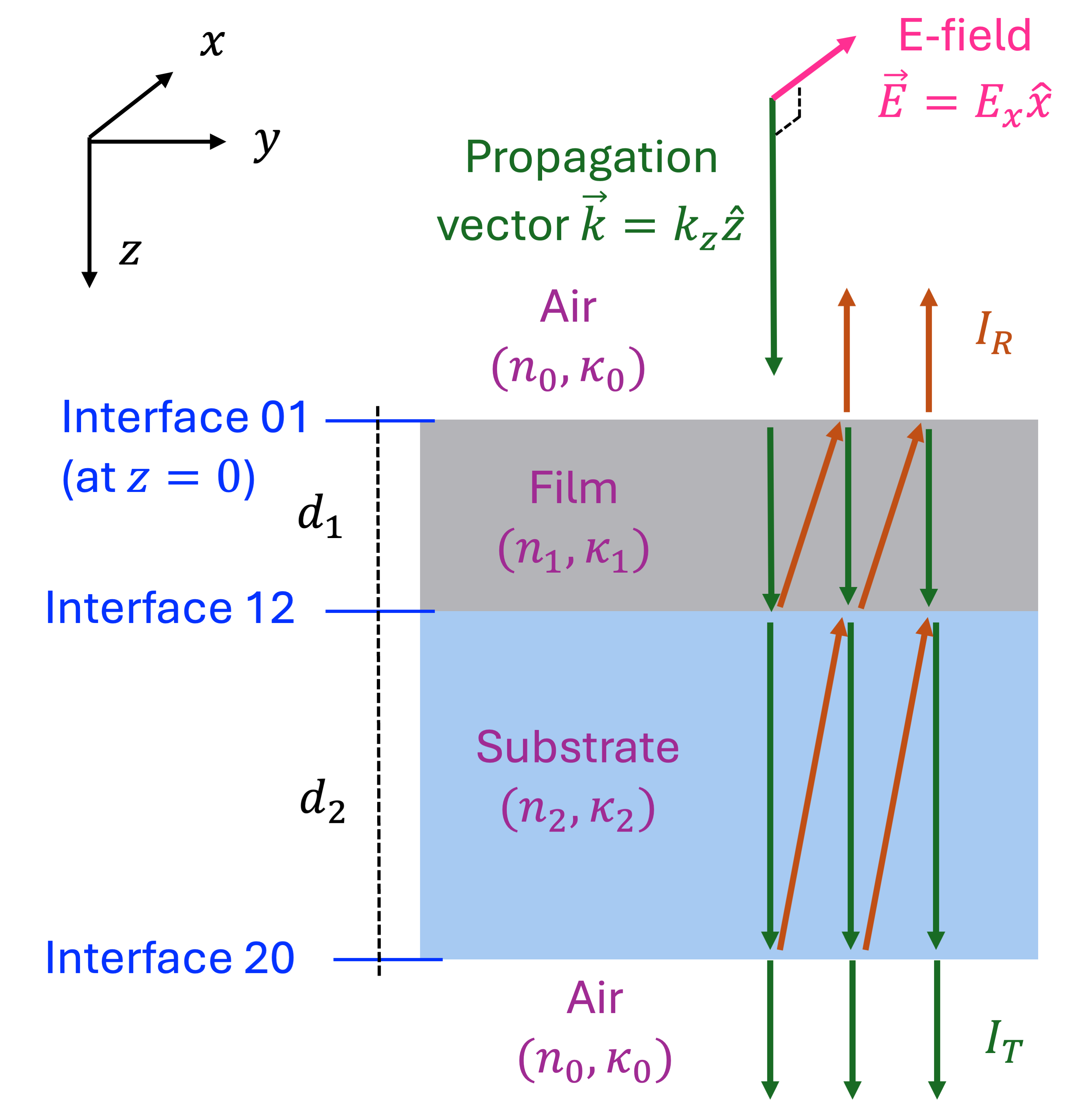}

   \caption{Scheme of the stratified media. Multiple interferences between incident and reflected waves occur because of the different interfaces.}
   \label{fig:model}
\end{figure}

Without loss of generality, let us consider a simple linearly-polarized light beam oriented in the $x$-direction, expressed as $\vec{E} = (E_x, 0, 0)$, perpendicular to the film's incidence plane (see Fig. \ref{fig:model}). Considering that this planar light wave propagates in the $z-$direction, the propagation vector becomes $\vec{k}=(0,0,2\pi/\lambda) = n (0,0,2\pi/\lambda_0) = n k_0 \hat{z}$, where $\lambda = \lambda_0/n$ is the wavelength in the material, $\lambda_0$ the wavelength in vacuum, and $k_0$ the wavenumber. Isotropic linear materials ensure the orthogonality between the propagation vector and the electric and magnetic fields \cite{born2013principles}, thus $\vec{H} = (0,H_y,0)$. 

If no external current is applied to the material $(J=0)$ and no free charges are induced in the bulk of the material $(\rho = 0)$, the Maxwell equations simplify as follows \cite{born2013principles}:
\begin{subequations}
\label{eq:main}
\begin{align}
\partial_x E_x &= 0 \label{eq:gauss_electric} \\
\partial_y H_y &= 0 \label{eq:gauss_magnetic} \\
\partial_z E_x &= \mu_0 \partial_t H_y \label{eq:faraday} \\
-\partial_z H_y &= \boldsymbol{\epsilon} \partial_t E_x \label{eq:ampere}
\end{align}
\label{eq:maxwell}
\end{subequations}

\subsection{Propagation matrix}
We will now focus only on propagation through layer 1 (the thin film). When the incident beam is a monochromatic continuous wave, the temporal dependency of the light wave becomes $e^{-i \omega t}$ \cite{born2013principles}, leading to an electric field of the form $E_x(x,y,z,t) = A(x,y,z) e^{-i \omega t}$. Here, $\omega=2\pi c/\lambda_0$ represents the angular frequency and $c$ the speed of light in vacuum. Incorporating this field expression into Eqs. \ref{eq:maxwell} and assuming constant material properties $\boldsymbol{\epsilon_1}$ and $\mu_0$ within the bulk of the layer, one can derive the so-called Helmholtz equation:
\begin{equation}
(\partial_{zz} + \boldsymbol{n}_1 k_0) A(x,y,z) = 0
\label{eq:helmholtz}
\end{equation}

\noindent A well-known solution \cite{born2013principles} for the amplitude $A(x,y,z)$ from Eq. \ref{eq:helmholtz} is a combination of two planar waves propagating in the $z-$ direction,
\begin{align}
    A(x,y,z) = A_\text{T}(z) + A_\text{R}(z) = \notag \\
    A_\text{0} \mathbf{t}_{02} e^{i {k_0} \mathbf{n_1} z} + A_\text{0} \mathbf{r}_{12} e^{- i {k_0} \mathbf{n_1} z} 
\label{eq:helmholtz_sol}
\end{align}
Eq. \ref{eq:helmholtz_sol} shows two distinct components, corresponding to a transmission wave (moving forward) and a reflection wave (moving backward) \cite{yeh1990optical}. Note that $\mathbf{t}_{01}$ and $\mathbf{r}_{12}$ represent the complex amplitude Fresnel coefficient for transmission and reflection, respectively. The transmission occurs between layer 0 and layer 1, while the reflection takes place from layer 2 toward layer 1. At normal incidence, these coefficients are defined as
\begin{subequations}
\begin{align}
    \mathbf{t}_{01} = \frac{2 \mathbf{n}_0}{\mathbf{n}_0 + \mathbf{n}_1} = \mathbf{r}_{01} + 1\label{eq:r_fresnel} \\
    \mathbf{r}_{12} = \frac{\mathbf{n}_1 - \mathbf{n}_2}{\mathbf{n}_1 + \mathbf{n}_2} =  \mathbf{t}_{12} 
 - 1\label{eq:t_fresnel}
\end{align}
\label{eq:fresnel}
\end{subequations}
Considering the amplitudes $A_\text{T}(z)$ and $A_\text{R}(z)$ immediately after the air-film interface (see Fig. \ref{fig:model}), our goal is to determine the transmitted and reflected wave amplitudes just before they encounter the next interface: the film-substrate boundary at $z + d_1$. By applying Eq. \ref{eq:helmholtz_sol} at $z + d_1$, we derive the following expressions using matrix formalism \cite{centurioni2005generalized}:
\begin{subequations}
\begin{align}
& \begin{pmatrix}
A_\text{T}(z+d_1) \\
A_\text{R}(z+d_1) 
\end{pmatrix} 
= 
P_1 \times 
\begin{pmatrix}
A_\text{T}(z) \\
A_\text{R}(z) 
\end{pmatrix} \label{eq:matrix_P} \\
& P_1 = \begin{pmatrix}
e^{i k_0 \mathbf{n_1} d_1} & 0 \\
0 & e^{-i k_0 \mathbf{n_1} d_1} 
\end{pmatrix}
\end{align}
\end{subequations}
The transfer matrix $P_1$ represents the \textit{propagation matrix} for layer 1. Note that the complex refractive index (located in the exponents of Eq. \ref{eq:matrix_P}) accounts for both phase delay and absorption through that homogeneous layer. Indeed, we can derive from Eq. \ref{eq:matrix_P} that a collimated light beam passing through the film during a single forward trip becomes
\begin{equation}
    A_\text{T}(z+d_1) = A_{T}(z) \exp{\bigg(-\frac{\alpha_1}{2} d_1\bigg)} \exp{\bigg(-\frac{i \delta_1}{2}\bigg)}
\label{eq:singletrip_trans}
\end{equation}
Where
\begin{subequations}
\begin{align}
    & \alpha_1 = \frac{4 \pi}{\lambda_0} \kappa_1 \label{eq:alpha_1}\\
    & \phi_1 = \frac{2 \pi}{\lambda_0} n_1 d \\
    & \delta_1 = 2 \phi_1
\end{align}
\label{eq:initial_var}
\end{subequations}

Here, $\phi_1$ refers to the phase delay introduced during the propagation of the beam through the film. It is more convenient to work instead with the variable $\delta_1$, which represents the phase delay during a whole round trip. Another relevant wavelength-dependent optical parameter is $\alpha_1$, known as the absorption coefficient. 

Considering the well-known relation \cite{hecht2012optics} between light intensity and wave amplitude, $I \propto |A|^2$, one can see from Eq. \ref{eq:singletrip_trans} that the transmitted intensity decreases exponentially with depth as $I_\text{T}(z+d_1) = I_\text{T}(z) e^{-\alpha_1 d_1}$, following the Beer-Lambert law \cite{hecht2012optics}. Note that $x_1 = I_\text{T}(z+d_1)/I_\text{T}(z) = e^{-\alpha_1 d}$ represents the transmittance corresponding to a single light trip. However, it should be noted that the multiple beam interferences within the two stratified media will lead to a different overall sample transmittance.

\begin{figure*}[b!]
  \centering
   \includegraphics[width=\linewidth]{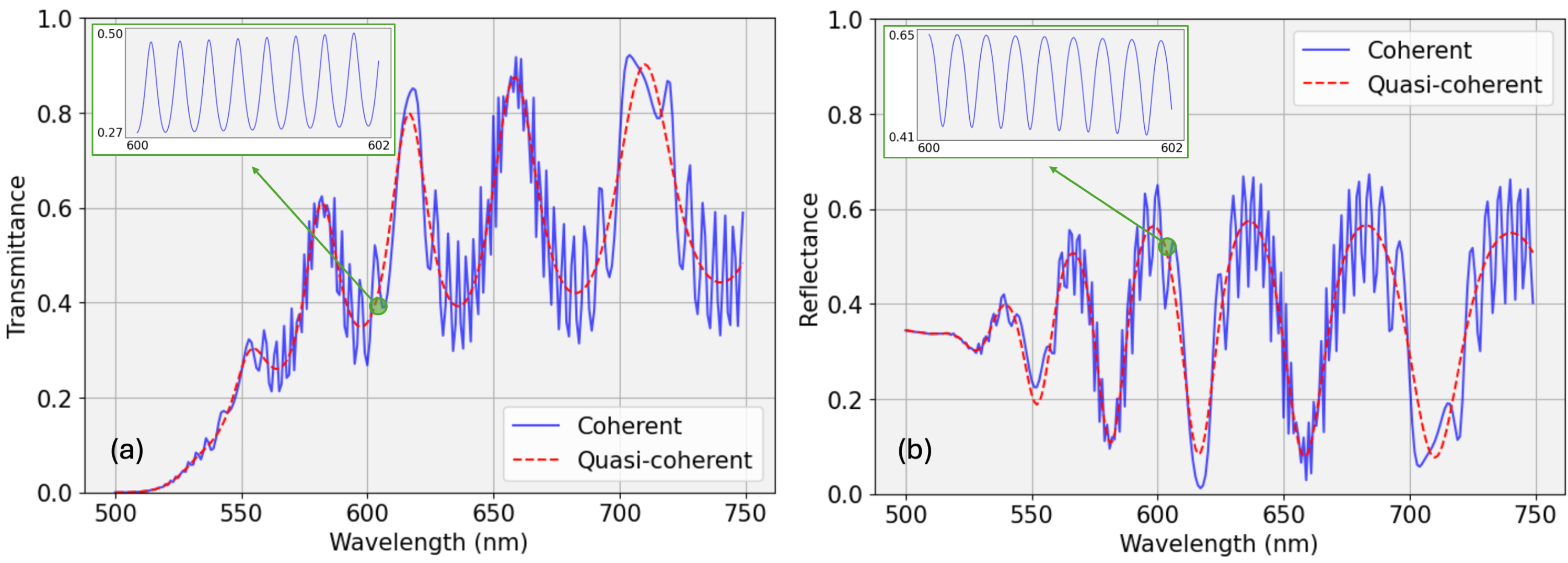}

   \caption{Transmittance (a) and reflectance (b) spectra of our simulated sample (for an amorphous silicon thin film), assuming a weakly absorbing glass substrate. The results are shown for a monochromatic light source with infinite coherence (Eqs. \ref{eq:coherence_T} and \ref{eq:coherence_R}) and for another light source with limited coherence (Eqs. \ref{eq:residue_T} and \ref{eq:residue_R}), where $2 d_1 < \ell < 2 d_2$.}
   \label{fig:coherent}
\end{figure*}

\subsection{Dynamic matrix}
Let us now consider the case where the beam reaches the interface between two layers, such as the thin film (layer 1) and the thick substrate (layer 2). 

Because the incident beam is normal to the sample surface, the electric and magnetic fields have only tangential components, represented as $(E_{1x}, H_{1y})$ for the film layer and $(E_{2x}, H_{2y})$ for the substrate layer. The tangent fields must remain continuous across the interface, meaning that $E_{1x} = E_{2x}$ and $H_{1y} = H_{2y}$ at $z = d$. Imposing these boundary conditions in Eqs. \ref{eq:maxwell} gives us the following relations \cite{yeh1990optical}:
\begin{subequations}
    \begin{align}
        A_\text{T}^{(1)} + A_\text{R}^{(1)} &= A_\text{T}^{(2)} + A_\text{R}^{(2)} \label{eq:cond1} \\
        \mathbf{n_1} (A_\text{T}^{(1)} - A_\text{R}^{(1)}) &= \mathbf{n_2} (A_\text{T}^{(2)} - A_\text{R}^{(2)}) \label{eq:cond2}
    \end{align}
    \label{eq:cond}
\end{subequations}

\noindent Here, $A_\text{T}^{(i)}$ and $A_\text{R}^{(i)}$ represent the amplitudes of the transmitted and reflected light waves within the layer $i\in\{1,2\}$, right next to the film-substrate interface. Eqs. \ref{eq:cond}, when transformed to matrix form, yield
\begin{equation}
\begin{pmatrix}
1 & 1 \\
\boldsymbol{n_1} & -\boldsymbol{n_1}
\end{pmatrix} 
\times 
\begin{pmatrix}
A_\text{T}^{(1)} \\
A_\text{R}^{(1)} 
\end{pmatrix} 
= 
\begin{pmatrix}
1 & 1 \\
\boldsymbol{n_2} & -\boldsymbol{n_2}
\end{pmatrix} 
\times
\begin{pmatrix}
A_\text{T}^{(2)} \\
A_\text{R}^{(2)} 
\end{pmatrix} 
\label{eq:matrix_D}
\end{equation}
After a few calculations, one can finally find
\begin{subequations}
\begin{align}
& \begin{pmatrix}
A_\text{T}^{(2)} \\
A_\text{R}^{(2)} 
\end{pmatrix} 
 = D_{12}
\times
\begin{pmatrix}
A_\text{T}^{(1)} \\
A_\text{R}^{(1)} 
\end{pmatrix} \label{eq:matrix_dynamic} \\
& D_{12} = \frac{1}{1+\mathbf{r_{12}}} \begin{pmatrix}
1 & -\mathbf{r_{12}} \\
-\mathbf{r_{12}} & 1
\end{pmatrix} 
\end{align}
\end{subequations}
The transfer matrix $D_{12}$ is called the \textit{dynamic matrix} and accounts for amplitude changes between the film-to-substrate (interface 12).

\subsection{Resulting transmittance and reflectance}

By sequentially computing the wave propagation through the layers and the interactions at the interfaces, the transmitted and reflected fields through the whole sample can be determined as:
\begin{equation}
\begin{pmatrix}
A_\text{T}^{(3)} \\
A_\text{R}^{(3)} 
\end{pmatrix} 
= 
M
\times
\begin{pmatrix}
A_\text{T}^{(0)} \\
A_\text{R}^{(0)} 
\end{pmatrix} 
\label{eq:matrix_total}
\end{equation}

\noindent Where the transfer matrix is now given by
\begin{equation}
M = D_{23}  \times P_2 \times D_{12} \times P_1 \times 
 D_{01}
\label{eq:matrix_M}
\end{equation}

\noindent Note that there is no light absorption throughout layers 0 and 3 (as it corresponds to air). Therefore, the propagation in these layers only leads to a constant phase delay, which will not affect the overall transmitted or reflected light intensity. The total transmittance and reflection can be calculated from the elements of the transfer matrix $M$, as explained in \cite{yeh1990optical}. We then obtain the following formulae:
\begin{subequations}
\begin{align}
    T = & \left| \frac{1}{M^{(1,1)}} \right|^2 = \frac{\mathbf{h}}{\boldsymbol{\tau} \boldsymbol{\bar{\tau}}} = \frac{h}{a + 2 b \cos \delta_2 + 2 c \sin \delta_2} \label{eq:coherence_T} \\
    R = & \left| \frac{M^{(1,2)}}{M^{(1,1)}} \right|^2 = \frac{\boldsymbol{\eta} \boldsymbol{\bar{\eta}}}{\boldsymbol{\tau} \boldsymbol{\bar{\tau}}} = \frac{e + 2 f \cos \delta_2 + 2 g \sin \delta_2}{a + 2 b \cos \delta_2 + 2 c \sin \delta_2} \label{eq:coherence_R}
\end{align}
\label{eq:matrix_TR}
\end{subequations}
Where the complex coefficients are defined as follows:
\begin{subequations}
{\small 
\begin{align}
    \mathbf{h} = & x_1 x_2 (\mathbf{r}_{01} + 1)(\mathbf{\bar{r}}_{01} + 1) \cdot \notag\\
    &(\mathbf{r}_{12} + 1)(\mathbf{\bar{r}}_{12} + 1)(\mathbf{r}_{23} + 1)(\mathbf{\bar{r}}_{23} + 1)\\
    \boldsymbol{\tau} = & \mathbf{r}_{01}(\mathbf{r}_{12} + \mathbf{r}_{23} x_2 e^{i \delta_2}) x_1 e^{i \delta_1} + (\mathbf{r}_{12} \mathbf{r}_{23} x_2 e^{i \delta_2} + 1) \\
    \boldsymbol{\eta} = & -\mathbf{r}_{01}(\mathbf{r}_{12} \mathbf{r}_{23} x_2 e^{i \delta_2} + 1) - (\mathbf{r}_{12} + \mathbf{r}_{23} x_2 e^{i \delta_2}) x_1 e^{i \delta_1}
\end{align} 
}
\label{eq:complex_coeff}
\end{subequations}

Note that the overall sample absorption can then be computed as $\mathcal{A} = 1 - (T + R)$ \cite{stadler2010analyzing}. The overbar on bold symbols, such as $\boldsymbol{\bar{\tau}}_2$, indicates the conjugate of complex numbers. Note that the sinusoidal functions from Eqs. \ref{eq:matrix_TR} depend on $\delta_2$ and account for the beam interferences within the substrate. Since the sample's transmittance and reflectance are \textit{scalar} wavelength-dependent functions, Eqs. \ref{eq:matrix_TR} can be equivalently written either in terms of complex coefficients $(\mathbf{h}, \boldsymbol{\tau}, \boldsymbol{\eta})$ or either in terms of real coefficients $(h,a,b,c,e,f,g)$, which are all defined in Appendix. Although Eqs. \ref{eq:matrix_TR} becomes more extensive when working with the real coefficients, it is much more convenient for computational purposes. It should also be clarified that the coefficients $(a, b, c)$, which appear in the denominator of Eqs. \ref{eq:matrix_TR}, contain sinusoidal functions that depend on the phase $\delta_1$. These sinusoidal functions account for the Fabry-Perot light beam interferences within the thin film.

Figure \ref{fig:coherent} (see curves in blue) shows the resulting transmission and reflection within the spectral range of 500 to 750 nm. In this work, we consider a simulated thin-layer sample that has a slightly absorbing glass substrate\footnote{In real samples, the absorption of the glass substrate varies with wavelength. For instance, the common Borofloat33 (a type of borosilicate substrate) exhibits significant absorption in the UV range (below 350 nm) and the NIR range (above 2100 nm) but is nearly transparent in the middle of the visible spectrum. Despite this quasi-transparency, subtle variations in transmission and reflection still occur. Researchers often limit their analysis to regions of evident transparency, though completely eliminating absorption effects is not feasible in practice (e.g., see the slight variations in substrate absorption lines in Fig. 3 of \cite{ballester2022application}). In our study, we use a simulated substrate with a constant weak absorption across the spectrum to provide a generalized analysis of errors in transmission and reflection formulas. This approach ensures that our findings are applicable to any type of substrate, as real substrates have diverse absorption curves.}, with fix values $n_2=1.5$, $\kappa_2=10^{-6}$, and $d_2=0.5$ mm. The optical properties of an amorphous silicon thin film, with thickness $d_1 = 1$ $\upmu$m, were simulated employing the empirical dispersion expressions $n_1 = 2.6 + 3 \cdot 10^5/\lambda^2$, $\log_{10}(\alpha_1) = -8 + 1.5 \cdot 10^6 / \lambda^2$ \cite{swanepoel1983determination}. Note that the extinction coefficient $\kappa_1$ can be directly determined from the absorption coefficient $\alpha_1$, by using Eq. \ref{eq:alpha_1}.

\section{Exact formulae for uniform films}
\label{sec:quasi-coherent-uniform}

So far, all the calculations have assumed a purely monochromatic light source, which introduces some unwanted high-frequency noise in the transmission and reflection spectra (see Figs. \ref{fig:coherent}). In the actual experiment, the light sources have a finite spectral bandwidth, which inherently limits the coherence length of the beam. By carefully selecting the bandwidth, we can avoid coherent interference within the substrate, effectively eliminating the associated noise.

For a standard light beam with Gaussian spectral shape, the coherence length $\ell$ can be determined from the specific central wavelength $\lambda_\text{c}$ and the bandwidth $\Delta \lambda$ (measured at full width half maximum). The established formula for this relationship is $\ell = 4 \ln(2)/\pi \lambda_\text{c}^2/\Delta \lambda$ \cite{akcay2002estimation}. The pure monochromatic light source ($\Delta \lambda \approx 0$) considered in Eqs. \ref{eq:matrix_TR} inherently led to an infinite coherence length. In this case, the interreflections both within the thin film and within the thick substrate resulted in the superposition of the electromagnetic waves, causing two distinct interference effects. 

According to the etalon interference condition \cite{hecht2012optics}, constructive interference occurs at wavelengths where $m = 2 n d / \lambda$ is an integer, leading to high values in the transmission spectra. Note that when $m$ is a half-integer, there is destructive interference, and we observe valleys in the spectra. According to this interference equation, the relatively small thickness of the film $d \approx 1 \,\upmu$m leads to low-frequency oscillations. These long oscillations appear as a characteristic sinusoidal pattern in the transmission spectrum, as illustrated in Figs. \ref{fig:coherent}a and \ref{fig:coherent}b (see dashed red lines). In contrast, the considerable thickness of the substrate, $d_\text{s} \approx 1$ mm, gives rise to high-frequency oscillations that can be interpreted as noise. 

The curves presented in Figs. \ref{fig:coherent}a and \ref{fig:coherent}b were plotted at each discrete wavelengths, 500, 501, 502, ..., and 750 nm. However, this spectral resolution is insufficient to clearly resolve the high-frequency oscillations originated by light wave interferences at the substrate, as illustrated in the insets of the figures (top left). Indeed, the insets show that there are eight oscillation peaks in a range of two nanometers. This limitation in sampling leads to the emergence of \textit{aliasing effects}. In spectral regions where the oscillation frequency is roughly a multiple of the \textit{sampling rate}, the curve appears to have fewer oscillations (e.g. 700 to 725 nm). Conversely, regions with greater misalignment exhibit rapid noisy oscillations (e.g. 725 to 750 nm).

Consequently, it is standard practice to deliberately adjust the bandwidth such that the coherence length $\ell$ is less than $2 \, d_\text{s}$, twice the thickness of the substrate. This adjustment mitigates the unwanted etalon effect from the glass, and it effectively removes the noise in the spectra, as the beam will become incoherent after a single round trip. Recall that the coherence length defines the distance over which the electromagnetic waves maintain their sinusoidal nature \cite{hecht2012optics}. When two coherent light beams with electric fields $E_1$ and $E_2$ overlap, they interfere linearly, resulting in a combined electric field, $E = E_1 + E_2$ \cite{hecht2012optics}. The resulting light intensity is given by $I = |E_1 + E_2|^2 = I_1 + I_2 + 2 E_1 E_2 \cos(\Delta theta_)$, where $\Delta theta$ is the phase difference between the two light beams. The interference term $2 E_1 E_2 \cos(\Delta theta_)$ oscillates rapidly with changes in wavelength. In contrast, for incoherent beams, the electric fields do not interfere, and the total intensity is simply $I =  I_1 + I_2$, with no oscillatory interference term.

\begin{figure*}[h!]
  \centering
   \includegraphics[width=\linewidth]{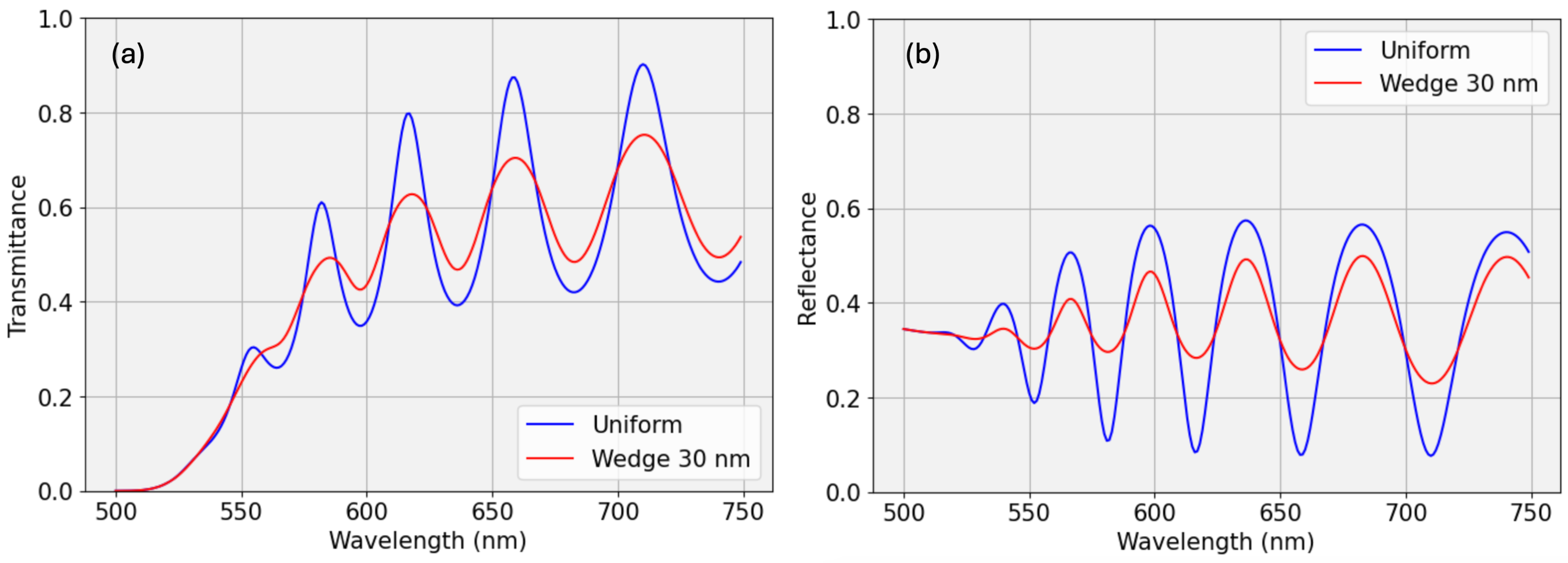}
   \caption{Sample (a) transmittance and (b) reflectance, using the exact formulae for uniform films (Eqs. \ref{eq:T_quasi}-\ref{eq:R_quasi}) and for wedge films (Eqs. \ref{eq:T_delta}-\ref{eq:R_delta}), with $\Delta d = 30$ nm. When the wedge increases, the contrast in the interferences decreases, as does the amplitude of the oscillations.}
   \label{fig:delta}
\end{figure*}

\begin{figure*}[b!]
  \centering
   \includegraphics[width=\linewidth]{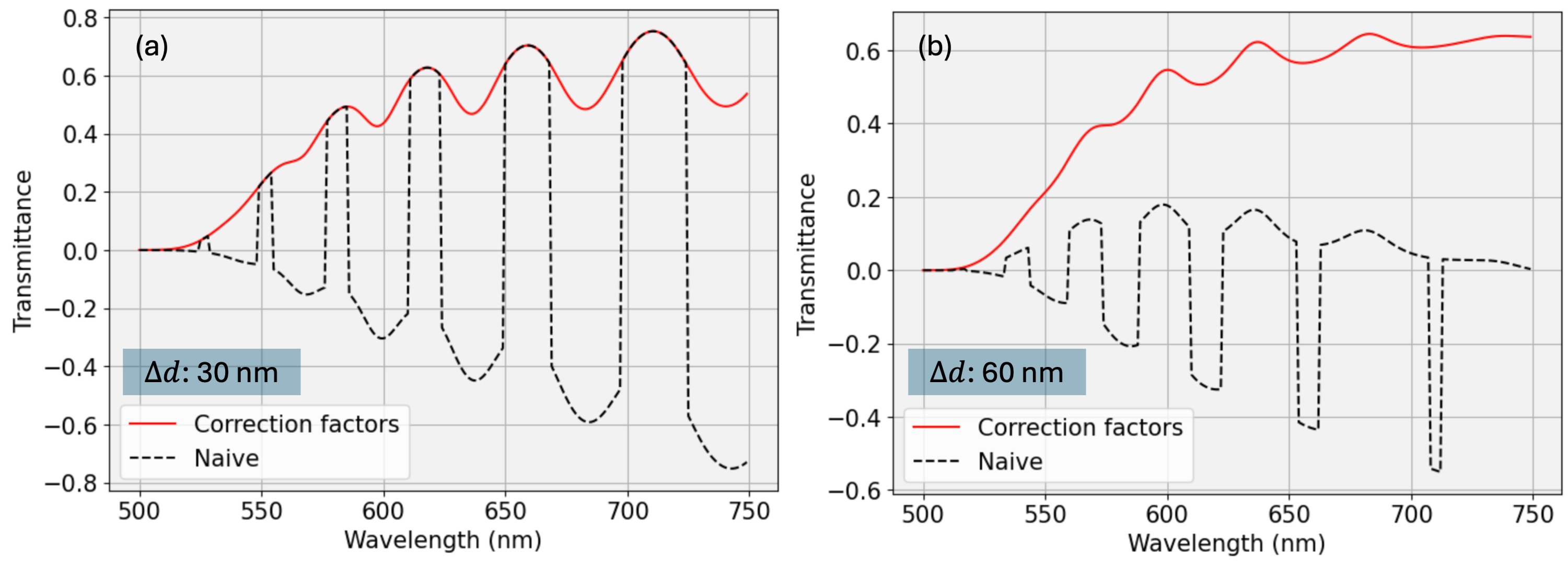}
   \caption{Simulated transmittance showing the effect of the correction factors $N^+$ and $N^-$ (see Eqs. \ref{eq:T_delta_all}) for thin films with two representative wedge parameters, (a) $\Delta d = 30$ nm and (b) $\Delta d = 60$ nm.}
   \label{fig:correction}
\end{figure*}

In many real experiments, the coherence length is set within $2 \, d < \ell < 2 \, d_\text{s}$. The multiple thin-film interreflections then result in the linear superposition of the electric fields (see Figs. \ref{fig:model}).  However, interreflections within the substrate lead to an incoherent superposition, producing just a simple sum of their intensities. To incorporate this complicated finite coherence model into our expressions from Eqs. \ref{eq:matrix_TR}, the \textit{spectral averaging} method \cite{yeh1990optical} proposes to average out the transmittance and reflectance over the corresponding phase delay introduced by the substrate, $\delta_2 = 4 \pi n_2 d_2 /\lambda$. For instance, consider an incident beam with a reasonable spectral bandwidth of 1 nm, center around the wavelength $\lambda_0 = 600$ nm. The spectrum of the light beam spans from $\lambda_0^- = 599.5$ nm to $\lambda_0^+ = 600.5$ nm, which leads to a phase delay change of $\Delta \delta_2 = |\delta_2^- - \delta_2^+| = 8 \pi$. The superposition of infinite waves with a phase delay range exceeding $2 \pi$ will effectively average out\footnote{Note that the phase delay change $\Delta \delta_2$ depends not only on the bandwidth but also on the central wavelength. As the central wavelength increases, the change in the phase delay decreases, making interference effects from the substrate more likely to appear in the sample's transmittance and reflectance. This can result in subtle noise at longer wavelengths. Most modern spectrophotometers \cite{marceaumeasurement} can mitigate this high-frequency noise, as well as photometric noise, using smoothing techniques, such as the Savitzky-Golay filter.}. Therefore, we can simply integrate the transmittance and reflectance from Eqs. \ref{eq:matrix_TR} within the substrate phase delay range $[0,2\pi]$, yielding
\begin{subequations}
{\small
\begin{align}
    T_\ell = \frac{1}{2 \pi} \int_0^{2 \pi} \frac{h \, \, \text{d} \delta_2}{a + 2 b \cos \delta_2 + 2 c \sin \delta_2} \\
    R_\ell = \frac{1}{2 \pi} \int_0^{2 \pi} \frac{(e + 2 f \cos \delta_2 + 2 g \sin \delta_2) \, \, \text{d} \delta_2}{a + 2 b \cos \delta_2 + 2 c \sin \delta_2}
\end{align}
}
\end{subequations}
Taking into account a change of variable, $\theta = e^{i \delta_2}$, we can conveniently calculate the phase-average reflectance and transmittance as
\begin{subequations}
{\small
\begin{align}
    T_\ell = \frac{1}{2 \pi} \oint_{|\theta |=1} \! \frac{h \, \, \text{d}\theta }{F(\theta )} \label{eq:T_quasi} \\
    R_\ell = \frac{1}{2 \pi} \oint_{|\theta|=1} \! \frac{G(z) \, \, \text{d}\theta }{F(\theta)} \label{eq:R_quasi} \\
    F(\theta) = \theta^2 (c + i b) + \theta (i a) + (-c + ib)\\
    G(\theta) = \theta^2 (g + i f) + \theta (i e) + (-g + if)
\end{align}
}
\end{subequations}
The denominator can now be factor as $F(\theta) = (c + ib) (\theta - \theta_1) (\theta - \theta_2)$, where 
\begin{subequations}
\begin{align}
    \theta_{1,2} = \frac{-a i \pm \sqrt{-a^2 + 4 (b^2 + c^2)}}{2 (c + ib)}, \label{eq:z12} \\
    |\theta_{1,2}| = \big|- \gamma \pm \sqrt{1 - \gamma^2} \big|, \, \, \, \gamma= \frac{a}{2 \sqrt{b^2 + c^2}} \in [0,1] \label{eq:z12_norm}
\end{align}
\end{subequations}
Note that the discriminant in Eq. \ref{eq:z12} is negative, as $a^2 \gg b^2, c^2$, making the numerator a pure imaginary number. This allows for a straightforward calculation of $|\theta_{1,2}|$ in Eq. \ref{eq:z12_norm}. It then becomes evident that $\theta_1$ (with a positive square root) lies within the unit disk of the complex plane, while $\theta_2$ lies beyond the disk. Applying the Cauchy Residue Theorem \cite{ablowitz2003complex} to Eq. \ref{eq:T_quasi}, we finally obtain the expression for the transmittance,
{\small
\begin{align}
    T_\ell = h \, i \, \mathcal{R}\bigg(\frac{1}{F(z)}, z_1 \bigg) = \frac{h \, i}{(c + i b)} \, \lim_{z \rightarrow z_2} \frac{(z - z_2)}{(z - z_1)(z-z_2)} = \notag \\
    \frac{h}{\sqrt{a^2 - 4 (b^2 + c^2)}} = \frac{h}{u} = \frac{h}{a' + 2 b' \cos \delta_1 + 2 c' \sin \delta_1} \label{eq:residue_T}
\end{align}
}
The variables $(u, a', b', c')$ are defined in Appendix. The coefficients $(a', b' , c')$ do not have any sinusoidal components. Therefore, the oscillations in transmittance $T_\ell$ come solely from the phase delay $\delta_1$ (see the sine and cosine in the denominator of Eq. \ref{eq:residue_T}). These oscillations are now only introduced by the thin film and not the substrate.

Employing an absolutely similar methodology, the analysis of reflectance results in the following formula:
\begin{equation}
    R_\ell = \frac{e}{u} - \frac{2 (bf + cg)}{u w}
\label{eq:residue_R}
\end{equation}
The expressions $(w, e, b, f, c, f)$ are also defined in the Appendix. Fig. 2 shows the transmission and reflection for a quasi-coherent beam. It can now be seen in that figure that the high-frequency noisy oscillations caused by light beam interference within the substrate have been correctly removed both in the transmission and reflection spectra.

\section{Novel formulae for strongly-wedged films}
\label{sec:quasi-coherent-wedge}

We now examine a film characterized by a wedge-shaped planar surface, denoted as $\Delta d > 0$. This wedge implies that the film thickness at the illuminated spot varies within the range $[d_1 - \Delta d, d_1 + \Delta d]$. In that case, the phase delay introduced by the film exhibits a variation from $\delta_1^- = 4 \pi n_1 (d_1 - \Delta d) / \lambda$ to $\delta_1^+ = 4 \pi n_1 (d_1 + \Delta d) / \lambda$, depending on the particular region within the illuminated spot. Considering a reasonable wedge $\Delta d \ll d_1$, the change in phase delay is typically $\Delta \delta_1 = |\delta_1^+ - \delta_1^-| \ll 2 \pi$. As a result, while the interference effects from the thin film are not completely eliminated, they can be significantly reduced depending on the value of $\Delta d$. A very large wedge (such as $\Delta d \approx d$) will eventually cancel out all interference effects. 

\begin{figure}[t!]
  \centering
   \includegraphics[width=0.8 \linewidth]{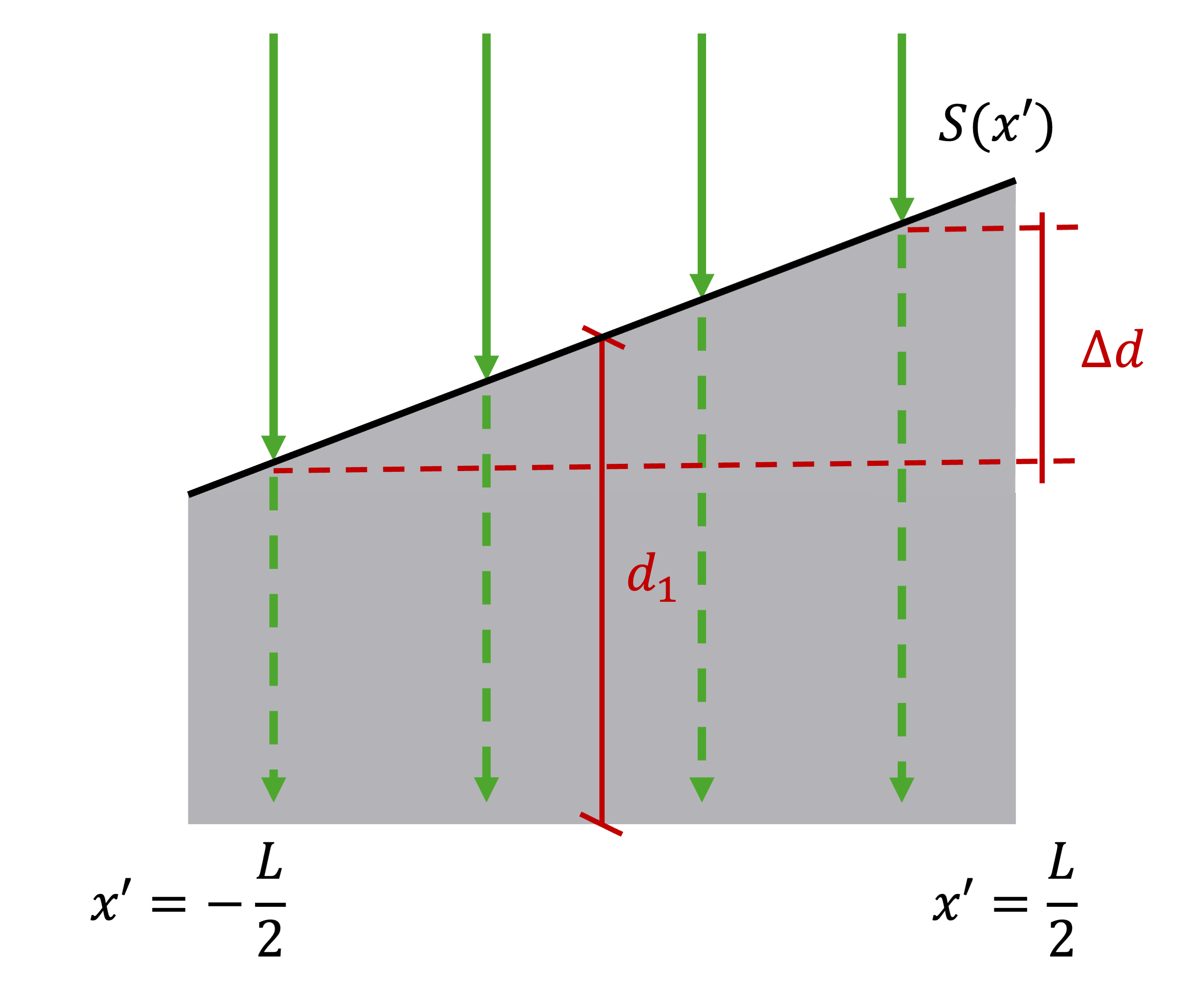}
   \caption{Profile of a tilted film with a linear surface described by $S(x') = mx'$. The optical path length in the $z-$direction varies depending on the position along the $x'-$axis.}
   \label{fig:intensity}
\end{figure}

Additionally, the film’s absorption will vary depending on the light path, whether it passes through $d_1 - \Delta d_1$, or $d_1 + \Delta d_1$, or somewhere in between (see Fig. \ref{fig:intensity}). Therefore, the single-trip transmittance $x_1$ should now be integrated from $x_1^- = e^{-\alpha_1 (d_1 - \Delta d)}$ to $x_1^+ = e^{-\alpha_1 (d_1 + \Delta d)}$. With the assistance of the symbolic software Wolfram Alpha Mathematica \cite{wolfra_alpha}, we were able to calculate the total transmittance of the sample,
\begin{subequations}
\small
\begin{align}
& T^\text{new}_{\Delta d} = \frac{1}{(\delta_1^+ - \delta_1^-) (x_1^+ - x_1^-)} \int^{x_1^+}_{x_1^-} \int^{\delta_1^+}_{\delta_1^-} \frac{h \, \, \text{d} \delta_1 \, \, \text{d} x_1}{a' + 2 b' \cos \delta_1 + 2 c' \sin \delta_1} \notag \\
&  \approx \frac{1}{\delta_1^+ - \delta_1^-} \int^{\delta_1^+}_{\delta_1^-} \frac{h \, \, \text{d} \delta_1}{a' + 2 b' \cos \delta_1 + 2 c' \sin \delta_1} = \label{eq:T_delta} \\
& \frac{2 \cdot h}{K \cdot (\delta_1^+ - \delta_1^-)} \left[ \arctan\left(\frac{I^+}{K}\right) + N^+ \pi - \arctan\left(\frac{I^-}{K}\right) - N^- \pi \right] \notag \\
& K = \sqrt{a'^2 - 4 \cdot (b'^2 + c'^2)} \\
& I^- = (a' - 2b') \tan\left(\frac{\delta_1^-}{2}\right) + 2c' \\ 
& I^+ = (a' - 2b') \tan\left(\frac{\delta_1^+}{2}\right) + 2c' \\
& N^+ = \text{round}\left(\frac{\delta_1^+}{2\pi}\right), \quad N^- = \text{round}\left(\frac{\delta_1^-}{2\pi}\right)
\end{align}
\label{eq:T_delta_all}
\end{subequations}
Note that in the intermediate steps, we applied the identity $\text{tanh}^{-1}(i \, q) = i \, \text{tan}^{-1}(q)$, valid for any value $q \in \mathbb{R}$. 

\begin{tcolorbox}[colback=yellow!10!white, colframe=red!50!black, title=Average light absorption in a wedge-shaped film, breakable] The average light absorption $x_1$ in a tilted film is approximately equivalent to that observed in a uniformly thick film, which is the reason for the approximation in Eq. \ref{eq:T_delta}. Although this assumption has been commonly used in various contexts \cite{ruiz2020optical}, it has not been previously supported by a rigorous mathematical proof. We will now explore the conditions under which this statement holds true. We introduce a local coordinate system $(x', y', z)$ such that the linear wedge is expressed solely in terms of the coordinate $x'$. The film profile can then be described as $S(x') = mx'$, where $m = 2 \Delta d/L$ represents the thin film slope. According to the Lambert-Beer law, the light intensity after the collimated beam passes through the film depth is given by $I(x') = I_0 e^{- \alpha_1 (d_1 + m x')}$. By integrating over the $x'-$range $[-L/2, L/2]$, we obtain
\begin{subequations}
\begin{align}
& \left\langle I_{\text{out}}(x') \right\rangle = \frac{1}{L} \int_{-\frac{L}{2}}^{\frac{L}{2}} I_0 e^{-\alpha_1 [d_1 + mx']} \, dx' \\
& = \frac{I_0}{L} e^{-\alpha_1 d_1} \left( \frac{-1}{\alpha_1 m} e^{-\alpha_1 mx'} \bigg|_{-\frac{L}{2}}^{\frac{L}{2}} \right) \\
& = I_0 e^{-\alpha_1 d_1} \frac{\sinh \left( \alpha_1 m L/2 \right)}{\alpha_1 m L/2}
\end{align}
\end{subequations}

We must take into account that $q = \alpha_1 m L/2 = \alpha_1 \Delta d$.  One should notice that $\sinh(q)/q \approx 1$ for $q \approx 0$, which justifies the aforementioned assumption for relatively small wedges, where $q$ remains close to zero. To quantify this, even in the case of a relatively large wedge $\Delta d = 60$ nm and high absorption $\alpha_1 = 10^{4}$ cm$^{-1}$, the wavelength-dependent term $\sinh(q)/q$ has an average value of 1.00060. Therefore, we can say that $I \approx I_0 e^{-\alpha_1 d_1}$ with an average error of around 0.06\%, which is negligible in practice, comparable to the typical photometric noise of the spectrophotometer. In these common scenarios, the single-pass light transmittance $x_1$ in a tilted film can be considered approximately equivalent to that observed in a uniform film.
\end{tcolorbox}

The film presents high absorption at 500 nm, as seen in Fig. \ref{fig:coherent}, where the transmittance eventually approaches zero.  At that wavelength, the term $\sinh(y)/y$ finds the highest value, 1.015. We used that approximation for the derivation of Eq. \ref{eq:T_delta}. and a maximum error of $1.5\%$ for the region of high absorption

Figs. \ref{fig:delta}a and \ref{fig:delta}b present the transmittance and reflectance simulated curves, respectively, for films with different wedges, $\Delta d = 0, 30, 60$ nm. We can see how higher wedges lead to less interference contrast, reducing the amplitude of the oscillations in the spectra.

Eqs. \ref{eq:T_delta_all} introduce the correction factors $N^+$ and $N^-$, which account for the different branches of the tangent function and facilitate the application of the formula to high wedges, where $\Delta d > \lambda/(4 n)$. These factors were previously motivated in \cite{ruiz2020optical} in the context of sample transmittance for non-absorbing substrates.

To clearly see the importance of the correction factors in the transmittance formula (see Eq. \ref{eq:T_delta_all}) when dealing with relatively high $\Delta d$ parameters, one can also see in Figs. \ref{fig:correction}a and \ref{fig:correction}b the plot of the spectra with and without the correction factors for the two representative cases $\Delta d = 30, 60$ nm. As the wedge increases, the naive approach (without correction factors) disagrees more with respect to the expected curves.

To the best of our knowledge, there is no equivalent \textit{closed-form} expression for the reflectance. Consequently, we propose using the trapezoidal rule \cite{atkinson1991introduction} for efficient and accurate numerical estimation:
\begin{align}
R^\text{new}_{\Delta d} = \frac{1}{\delta_1^+ - \delta_1^-} \int_{\delta_1^+}^{\delta_1^+} \bigg( \frac{e}{u} - \frac{2 (bf + cg)}{u w} \bigg)  \, d\delta_1 = \notag \\
 \frac{1}{2} \bigg[ R_L \bigg(\delta_1^- + 2 \Delta d \frac{0}{N} \bigg) + 2 R_L\bigg(\delta_1^- + 2 \Delta d \frac{1}{N}\bigg) + ... \label{eq:R_delta} \\
 ... + 2 R_L\bigg(\delta_1^- + 2 \Delta d \frac{N-1}{N}\bigg) + R_L\bigg(\delta_1^- + 2 \Delta d \frac{N}{N}\bigg) \bigg] \notag
\end{align}
Where $R_L(\delta_1)$ is the reflectance for a uniform film as a function of the phase $\delta_1$, as given by Eq. \ref{eq:residue_R}. Our findings indicate that setting $N = 40$ points for the phase already yields reasonable results, since we obtain a total absorptance value $\mathcal{A}$ effectively zero when \textit{no} film or substrate absorption is added ($\kappa_1 = \kappa_2 = 0$),  $\mathcal{A} = |1 - T_{\Delta d} - R_{\Delta d}| < 10^{-2}$. In any case, note that we can define a sufficiently fine sampling that meets any desired level of accuracy.

\begin{table*}[t!]
\centering
\small
\begin{tabular}{|>{\columncolor{lightgray}}c|>{\columncolor{lightgray}}c|c|c|c|c|}
\hline
\textbf{$\boldsymbol{\kappa_2}$} & \textbf{$\boldsymbol{\Delta d}$ (nm)} & \multicolumn{1}{|m{3cm}|}{\centering \textbf{Transmission (RP)} \\ \textbf{Eqs. \ref{eq:T_JJ_total}} \\ \textbf{RMSE (\%)}} & \multicolumn{1}{|m{3cm}|}{\centering \textbf{Reflection (Minkov)} \\ \textbf{Eqs. \ref{eq:Minkov_total}} \\ \textbf{RMSE (\%)}} & \multicolumn{1}{|m{3.7cm}|}{\centering \textbf{Transmission (Swanepoel)} \\ \textbf{Eqs. \ref{eq:Swan83_total}-\ref{eq:Swan84_total}} \\ \textbf{RMSE (\%)}} & \multicolumn{1}{|m{3.5cm}|}{\centering \textbf{Reflection (Swanepoel)} \\ \textbf{Eqs. \ref{eq:swanR_total}-\ref{eq:swanRdelta_total}} \\ \textbf{RMSE (\%)}} \\
\hline
0 & 0 & 0 & $10^{-4}$ & 0.035 & 0.068 \\
\hline
0 & $10^{-5}$ & 0 & - & 0.035 & 0.068 \\
\hline
0 & 30 & 0 & - & 0.012 & 0.067 \\
\hline
0 & 60 & 0 & - & 0.007 & 0.067 \\
\hline
$10^{-6}$ & 0 & 0.496 & 0.034 & 0.497 & 0.074 \\
\hline
$10^{-6}$ & $10^{-5}$ & 0.496 & - & 0.497 & 0.074 \\
\hline
$10^{-6}$ & 30 & 0.487 & - & 0.487 & 0.073 \\
\hline
$10^{-6}$ & 60 & 0.488 & - & 0.488 & 0.073 \\
\hline
\end{tabular}
\caption{Errors associated with the approximated formulae. When $\Delta d = 0$ the formulae for uniform films are used, tested against the exact formulae from Eqs. \ref{eq:T_quasi} and \ref{eq:R_quasi}. For $\Delta d = 10^{-5}$ nm, the formulae for wedged films are applied, demonstrating consistency as $\Delta d \rightarrow 0$. They were tested against the new formulae from Eqs. \ref{eq:T_delta} and \ref{eq:R_delta}. Note that the two left blocks assume $\kappa_2 = 0$, while the two right blocks use the Swanepoel approximations ($\kappa_2 = 0$ and $n_1^2 \ll k_1^2$).}
\label{tab:approx}
\end{table*}

\section{Particular case: Non-absorbing substrate} 

\subsection{Transmittance}
The newly derived transmittance formula, Eq. \ref{eq:T_delta}, closely matches that found by Ruiz-Perez (RP) \textit{et al.} in \cite{ruiz2020optical}. However, our expression now incorporates the possibility of substrate absorption. Indeed, the RP equation (without substrate absorption) can be alternatively derived now\footnote{We found an errata in the RP formula presented in  \cite{ruiz2020optical}. In particular, their Eq. 6 incorrectly omits a minus sign before the term ``$C_{22} x \sin(\phi)$''. This error can be confirmed by comparing it with Eq. A1 in \cite{swanepoel1983determination}. Consequently, Eqs. 14 in \cite{ruiz2020optical}, corresponding to the formulae for the coefficients ``$K_1$'' and ``$K_2$'', should also include a minus sign before ``$C_{22} x$''.} by simply setting $\kappa_2 = 0$ in our new Eq. \ref{eq:T_delta}. Following the notation from RP, we then get the following simplified formulae
\begin{subequations}
\small
\begin{align}
& A = 16 (n_1^2 + k_1^2) n_2, \\
& B = ((n_1 + 1)^2 + k_1^2)((n_1 + 1)(n_1 + n_2^2) + k_1^2), \\
& C_1 = 2\left((n_1^2 + k_1^2 - 1)(n_1^2 + k_1^2 - n_2^2) - 2 k_1^2 (n_2^2 + 1)\right), \\
& C_2 = 2 k_1 \left(2(n_1^2 + k_1^2 - n_2^2) + (n_1^2 + k_1^2 - 1)(n_2^2 + 1)\right), \\
& D = ((n_1 - 1)(n_1 - n_2^2) + k_1^2)((n_1 - 1)^2 + k_1^2), \\
& T^\text{RP20} = \frac{A \, x_1}{B - C_1 \, x_1 \, \cos(\delta_1) + C_2 \, x_1 \, \sin(\delta_1) + D \, x_1^2} \label{eq:T_JJ_uniform} \\
& F = (B + D x_1^2 + C_1 x_1) \tan\left(\frac{\delta_1^-}{2}\right) + C_2 x_1, \\
& G = (B + D x_1^2 + C_1 x_1) \tan\left(\frac{\delta_1^+}{2}\right) + C_2 x_1, \\
& H = B^2 - x_1^2 (C_1^2 + C_2^2 - 2BD - D^2 x_1^2), \\
& T^\text{RP20}_{\Delta d} = \frac{2 A x_1}{(\delta_1^+ - \delta_1^-) \sqrt{H}} \left[\arctan\left(\frac{G}{\sqrt{H}}\right) + N^+ \pi - \right. \label{eq:T_JJ} \\
& \quad \quad \quad \quad \left. \arctan\left(\frac{F}{\sqrt{H}}\right) - N^- \pi \right] \notag
\end{align}
\label{eq:T_JJ_total}
\end{subequations}
Note that $T^\text{RP20}$ from Eq. \ref{eq:T_JJ_uniform} corresponds to the transmittance of the sample with a uniform film thickness, without taking into account substrate absorption. Similarly,  $T^\text{RP20}_{\Delta d}$ from Eq. \ref{eq:T_JJ} represents the case in which the sample has a certain wedge $\Delta d > 0$. We have confirmed numerically that, when no substrate absorption is considered (i.e., $\kappa_2 = 0$), the root mean square error (RMSE) is zero for any wedge parameter between \textit{(i)} our new exact formula from Eq. \ref{eq:T_delta}, and \textit{(ii)} the approximate expression from Eq. \ref{eq:T_JJ}. When $\kappa_2 = 10^{-6}$ and $\Delta d = 30$ nm, the RMSE of the approximated formulae by RP goes up to 0.465\%. This finding clearly shows the importance of incorporating substrate absorption into the models. More results for different wedges are summarized in Table \ref{tab:approx}. One can see that, as the wedge increases, the error decreases. This happens because the smaller the oscillation amplitude, the smoother the curve, getting lower peak values. Similarly, we have checked for the case of uniform films ($\Delta d = 0$) that the exact formula Eq. \ref{eq:T_quasi} and Eq. \ref{eq:T_JJ_uniform} produce exactly the same numerical results when no substrate absorption is considered.

\subsection{Reflectance}
Based on the foundational work of Grebenstikov \textit{et al.} \cite{grebenstikov1946decreasing}, Minkov \textit{et al.} \cite{minkov1989calculation} derived in 1989 an accurate expression for reflectance (which excludes substrate absorption). While Minkov's equation is only applies to films with uniform thickness, an analytical expression for wedge-shaped films can be simply derived if we further assume $\kappa_1 \ll n_1$ (as will be explaine in detail in the next section).

One can observe that Minkov's formula, Eq. \ref{eq:R_Minkov89}, is \textit{not} completely equivalent to the exact expression derived from Eq. \ref{eq:residue_R} when substrate absorption is disregarded ($\kappa_2 = 0$). However, the numerical discrepancy between them is minimal. For instance, we found a value of discrepancy error as low as $10^{-4}$\% when using our simulated sample. The discrepancy arises because, in addition to the assumption that $\kappa_2 = 0$, second-order approximations involving $n_1$ and $\kappa_1$ were also utilized in Minkov's expression.

However, when considering the weakly absorbing substrate from our simulated sample, the RMSE of Minkov's formula rises up to 0.034\%. It should be highlighted that the reflection formula, Eq. \ref{eq:R_Minkov89}, is considerably less affected by the substrate absorption than the transmission formula from Eq. \ref{eq:T_JJ} (see Table \ref{tab:approx}): As mentioned earlier, the overall reflectance and transmittance can be expressed as the addition of infinite light wave components, which appear due to the inter-reflections at each interface (see Fig. \ref{fig:model}). When analyzing the overall reflectance, the two \textit{main} components come from the first air-film interface and the second film-glass interface. The third component from the last substrate-glass interface (thus affected by the substrate absorption) is certainly weaker. In contrast, all transmission components account for the beam passing through the whole substrate volume. Therefore, transmittance measurements are notably affected by the existing substrate absorption.

The main advantage of Minkov's reflectance formula, and the reason of its popularity \cite{humphrey2007direct, luvnavcek2009simple}, is that it provides a simpler and more accessible expression than simply setting $\kappa_2 = 0$ in our new complicated formula, Eq. \ref{eq:residue_R}. Additionally, Minkov's formula facilitates the analysis of reflection for the envelope method, as explored in the next section. 

Minkov's formula is shown below:

\begin{subequations}
\small
\begin{align}
& A' = \left((n_1 - 1)^2 + k_1^2\right) \left((n_1 + n_2)^2 + k_1^2\right) \\
& B_1' = 2 \left((n_1^2 + k_1^2 - 1)(n_1^2 + k_1^2 - n_2^2) + 4 k_1^2 n_2\right) \\
& B_2' = 4 k_1 \left(n_2 (n_1^2 + k_1^2 - 1) - (n_1^2 + k_1^2 - n_2^2)\right) \\
& C' = \left((n_1 + 1)^2 + k_1^2\right) \left((n_1 - n_2)^2 + k_1^2\right) \\
& A'' = \left((n_1 + 1)^2 + k_1^2\right) \left((n_1 + n_2)^2 + k_1^2\right) \\
& B_1'' = 2 \left((n_1^2 + k_1^2 - 1) (n_1^2 + k_1^2 - n_2^2) - 4 k_1^2 n_2\right) \\
& B_2'' = 4 k_1 \left(n_2 (n_1^2 + k_1^2 - 1) + (n_1^2 + k_1^2 - n_2^2)\right) \\
& C'' = \left((n_1 - 1)^2 + k_1^2\right) \left((n_1 - n_2)^2 + k_1^2\right) \\
& G' = 64 n_2 (n_2 - 1)^2 (n_1^2 + k_1^2)^2 \\
& D'' = \left((n_1 + 1)^2 + k_1^2\right) \left((n_1 + 1)(n_1 + n_2^2) + k_1^2\right) \\
& E_1'' = 2 \left((n_1^2 + k_1^2 - 1) (n_1^2 + k_1^2 - n_2^2) - 2 k_1^2 (n_2^2 + 1)\right) \\
& E_2'' = 2 k_1 \left((n_1^2 + k_1^2 - n_2^2) + (n_2^2 + 1) (n_1^2 + k_1^2 - 1)\right) \\
& F'' = \left((n_1 - 1)^2 + k_1^2\right) \left((n_1 - 1)(n_1 - n_2^2) + k_1^2\right) \\
& R^\text{Mink89} = \frac{A' - (B_1' \cos(\delta_1) - B_2' \sin(\delta_1)) x_1 + C' x_1^2}{A'' - (B_1'' \cos(\delta_1) - B_2'' \sin(\delta_1)) x_1 + C'' x_1^2} + \notag \\ 
& \quad \quad \frac{G' \, x_1^2}{A'' - (B_1'' \cos(\delta_1) - B_2'' \sin(\delta_1)) x_1 + C'' x_1^2} \times \label{eq:R_Minkov89} \\
& \quad \quad \frac{1}{D'' - (E_1'' \cos(\delta_1) - E_2'' \sin(\delta_1)) x_1 + F'' x_1^2} \notag
\end{align}
\label{eq:Minkov_total}
\end{subequations}

\section{The Swanepoel Approximations}
\label{sec:envelopes}

\begin{figure*}[t!]
  \centering
   \includegraphics[width=\linewidth]{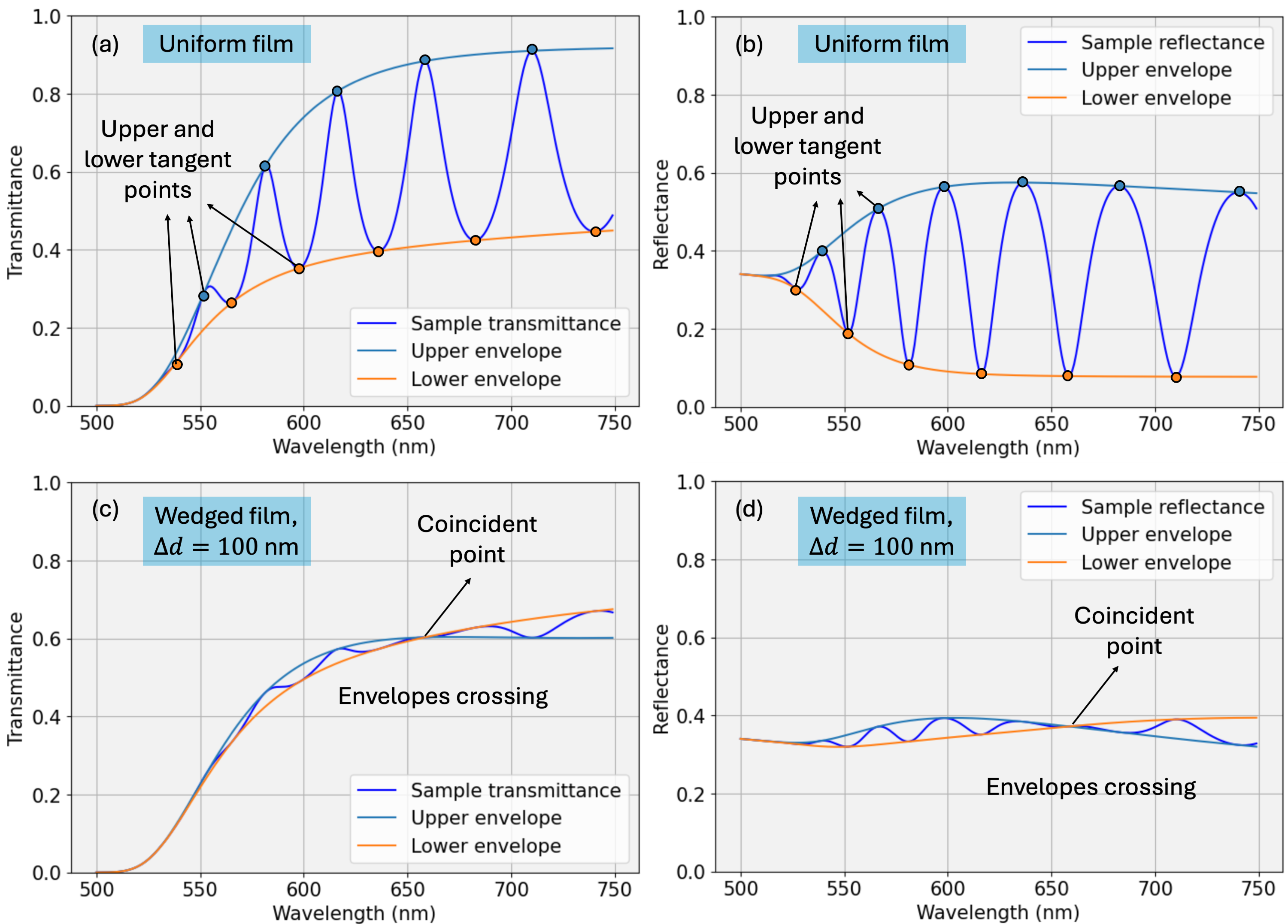}
   \caption{Sample transmittance (left column) and reflectance (right column) using the Swanepoel approximations. The upper and lower envelopes, as well as the tangent points where the spectrum intersects the envelopes, are shown. (a) and (b) represent uniform films (see Eqs. \ref{eq:Swan83_total} and \ref{eq:swanR_total}), while (c) and (d) correspond to films with a very high wedge, $\Delta d = 100$ nm (see Eqs. \ref{eq:Swan84_total} and \ref{eq:swanRdelta_total}). Under high wedge condition, we see how the envelopes \textit{cross over}, reaching a particular wavelength in which the spectrum and the two envelopes coincide.}
   \label{fig:swanepoel}
\end{figure*}

To ensure the completeness of this work, and to facilitate comparison with our results, we will now examine other previously established formulae for $T$ and $R$ that have been widely employed in the literature. As mentioned before, Swanepoel refined an algebraic procedure, commonly known as the envelope method, to directly determine $(n_1, \kappa_1, d)$. However, this method finds the optical properties only at specific wavelengths $\lambda_i$ corresponding to the maxima or minima Fabry-Perot interferences. In particular, these critical points are situated near the peaks and valleys of the transmission spectra (see Figs. 5a and 5b). In practice, they are identified by the intersection of the experimental spectra with their respective envelopes (reason why these intersection points are often described as the ``tangent points'').

This envelope method operates under two assumptions: \textit{(i)} the substrate does not absorb ($\kappa_2 = 0$), and (2) the extinction coefficient of the film is much weaker than the refractive index $\kappa_1^2 \ll n_1^2$. These two conditions are known as the Swanepoel approximations. Note that we analyzed condition \textit{(i)} only in Section 7. Condition \textit{(ii)} is typically valid for spectral regions with medium-to-weak absorption, where the transmittance alone indeed contains sufficient information for characterization. This second assumption, however, does not hold in areas of strong absorption, particularly near the optical band gap, where the transmission $T$ clearly decreases, eventually approaching and reaching zero.

\subsection{Transmission}
By setting $\kappa_2 = 0$ in Eq. \ref{eq:residue_T}, the transmission through a sample with a uniform-thickness film simplifies \cite{swanepoel1983determination} to the following expression:
\begin{subequations}
\begin{align}
& A_0 = 16 n_1^2 n_2 \\
& B_0 = (n_1 + 1)^2 (n_1 + 1) (n_1 + n_2^2) \\
& C_0 = 2 (n_1^2 - 1) (n_1^2 - n_2^2) \\
& D_0 = (n_1 - 1)^3 (n_1 - n_2^2) \\
& T^\text{Swan83} = \frac{A_0 x_1}{B_0 - C_0 x_1 \cos(\delta_1) + D_0 x_1^2} \label{eq:T_swan83} \\
& T_\text{M}^\text{Swan83} = \frac{A_0 x_1}{B_0 - C_0 x_1 + D_0 x_1^2} \label{eq:T_swan83_M} \\
& T_\text{m}^\text{Swan83} = \frac{A_0 x_1}{B_0 + C_0 x_1 + D_0 x_1^2} \label{eq:T_swan83_m}
\end{align}
\label{eq:Swan83_total}
\end{subequations}
It must be pointed out that the upper and lower envelopes of the spectrum from Eqs. \ref{eq:T_swan83_M} and \ref{eq:T_swan83_m} are found by fixing the sinusoidal component of the transmittance in Eq. \ref{eq:T_swan83} to the maximum or minimum value, that is, $\cos(\delta_1) = \pm 1$.

The set of expressions from Eq. \ref{eq:Swan83_total} were first derived in the Swanepoel seminal paper from 1983 \cite{swanepoel1983determination}. When no substrate absorption is considered ($\kappa_2=0$), the RMSE between our exact Eq. \ref{eq:T_delta} and the approximated Eq. \ref{eq:T_swan83} (which assumes $\kappa_1^2 \ll n_1^2$) is 0.039\% for our simulated thin-film sample. When substrate absorption is considered, the error increases to 0.472\%.

For non-uniform thin films with a certain wedge parameter $\Delta d>0$, the expressions for the transmission and envelopes are rewritten as follows:
\begin{subequations}
{\small
\begin{align}
    &F_0 = \frac{A_0 x_1}{B_0 + D_0 x_1^2}, \quad G_0 = \frac{C_0 x_1}{B_0 + D_0 x_1^2} \\
    &I_0^+ = \frac{1 + G_0}{\sqrt{1 - G_0^2}} \tan\left(\frac{\delta_1^+}{2}\right) \\ 
    &I_0^- = \frac{1 + G_0}{\sqrt{1 - G_0^2}} \tan\left(\frac{\delta_1^-}{2}\right) \\
    &T^\text{Swan84}_{\Delta d} = \frac{\lambda}{4 \pi n_1 \Delta d} \frac{F_0}{\sqrt{1 - G_0^2}} \bigg[ \arctan(I_0^+) + N^+ \pi - \notag \\ 
    & \quad \quad \quad \quad \quad \arctan(I_0^-) - N^- \pi \bigg] \label{eq:T_swan84} \\ 
    &I_\text{M} = \frac{1 + G_0}{\sqrt{1 - G_0^2}} \tan\left(\frac{2 \pi n_1 \Delta d}{\lambda}\right) \\
    &I_\text{m} = \frac{1 - G_0}{\sqrt{1 - G_0^2}} \tan\left(\frac{2 \pi n_1 \Delta d}{\lambda}\right) \\
    & N_\delta = \text{round}\bigg( \frac{\delta_1^+ - \delta_1^+}{2 \cdot 2 \pi} \bigg)\\
    &T^\text{Swan84}_\text{M}  = \frac{\lambda}{2 \pi n_1 \Delta d} \frac{F_0}{\sqrt{1 - G_0^2}} ( \arctan(I_\text{M}) + N_\delta \, \pi) \\
    &T^\text{Swan84}_\text{m} = \frac{\lambda}{2 \pi n_1 \Delta d} \frac{F_0}{\sqrt{1 - G_0^2}} ( \arctan(I_\text{m}) + N_\delta \, \pi)
\end{align}
}
\label{eq:Swan84_total}
\end{subequations}
These formulae for wedged-shaped films were initially found in the subsequent seminal paper by Swanepoel from 1984 \cite{swanepoel1984determination}. Please consider that we have also added the correction numbers $N^+$ and $N^-$ to account for highly tilted films, $\Delta d > \lambda/(4 n)$, and we have also included a correction factor $N_\delta$ for the envelopes. An interesting detail is that when $\Delta d > \lambda/(4 n)$ holds, the upper and lower envelopes cross: the lower envelope will stay above the spectra and the upper envelope below (see Fig. 5c). Over a wide spectral range, one can see how the two envelopes alternate sequentially. Therefore, there are particular wavelengths at which the spectra $T^\text{Swan84}_{\Delta d}$ and its two envelopes precisely coincide, as seen in Figs. \ref{fig:swanepoel}c-d. The crossover points are located at $\lambda_\text{cross} = 4 n \Delta d/N$, for $N = 1, 2, 3, ...$, as discussed in \cite{ruiz2020optical}. These particular points contain essential information and allow us to extract accurate information for the optical properties. 

The RMSE between the exact Eq. \ref{eq:T_delta} and the approximated Eq. \ref{eq:T_swan84} is 0.465\% when substrate absorption is taken into considering and $\Delta d = 30$ nm. The error from this formula is roughly equivalent to the error from the RP formula in Eq. \ref{eq:T_JJ}. The RP formula (which removes the assumption $\kappa_1^2 \ll n_1^2$) offers improvements only up to the fourth decimal place.

Setting $\kappa_2 = 0$ and $\Delta d = 10^{-5}$ nm in Eq. \ref{eq:T_swan84} leads to a transmission error of 0.039\%. As expected, the result obtained was identical to that previously obtained by using Eq. \ref{eq:T_swan83} for uniform films. This consistency occurs because Eq. \ref{eq:T_swan84} converges to Eq. \ref{eq:T_swan83} as $\Delta d \rightarrow 0$.

\subsection{Reflection}
Setting $\kappa_1^2 \ll n_1^2$ in Minkov's reflection formula (see Eq. \ref{eq:R_Minkov89}), we derive the following simplified expression for uniform films:
\begin{subequations}
{\small
\begin{align}
    &a_0 = n_1 - 1, \, \, \, \, \quad \quad  b_0 = n_1 + 1 \\
    &c_0 = n_1 - n_2, \quad \quad d_0 = n_1 + n_2 \\
    &e_0 = n_1 - n_2^2, \quad \quad f_0 = n_1 + n_2^2 \\
    &g_0 = 64 \, n_2 \, (n_2 - 1)^2 \, n_1^4 \\
    &R^\text{RP01} = \frac{(a_0 d_0)^2 + (b_0 c_0 x_1)^2 - 2 a_0 b_0 c_0 d_0 x_1 \cos(\delta_1)}{(b_0 d_0)^2 + (a_0 c_0 x_1)^2 - 2 a_0 b_0 c_0 d_0 x_1 \cos(\delta_1)} + \notag\\
    & \quad \quad \frac{g_0 x_1^2}{(b_0 d_0)^2 + (a_0 c_0 x_1)^2 - 2 a_0 b_0 c_0 d_0 x_1 \cos(\delta_1)} \times \label{eq:R_RP01_swan} \\
    & \quad \quad \frac{1}{b_0^3 f_0 + a_0^3 e_0 x_1^2 - 2 a_0 b_0 c_0 d_0 x_1 \cos(\delta_1)} \notag \\
    &R^\text{RP01}_\text{M/m} = \frac{(a_0 d_0 \pm b_0 c_0 x_1)^2}{(b_0 d_0 \pm a_0 c_0 x_1)^2} + \frac{g_0 x_1^2}{(b_0 d_0 \pm a_0 c_0 x_1)^2} \times \label{eq:R_RP01_swan_M} \\
    & \quad \quad \quad \frac{1}{\left(b_0^3 f_0 + a_0^3 e_0 x_1^2 \pm 2 a_0 b_0 c_0 d_0 x_1\right)} \notag
\end{align}
}
\label{eq:swanR_total}
\end{subequations}
We have verified that these results precisely match the exact formulae from Eq. \ref{eq:residue_R} when $\kappa_2 = 0$ and $\kappa_1^2 \ll n_1^2$. Note that the way to effectively set this last approximation is to write $\kappa_1^2 = 0$ when that term is compared to $n_1^2$ within summations. The upper and lower envelopes defined in Eq. \ref{eq:R_RP01_swan_M} correspond to the positive ($+$) and negative ($-$) signs, respectively.  These formulae were partially developed by Minkov \textit{et al.} \cite{minkov1989calculation} in 1989.  The RMSE of Eq. \ref{eq:R_RP01_swan} is 0.42\% for absorbing substrates and 0.02\% for transparent substrates.

The work \cite{ruiz2001method} by RP \textit{et al.} from 2001 also incorporates equations that address the scenario of weakly-wedge shaped films:
\begin{subequations}
{\small
\begin{align}
    & L_0 = b_0^2  d_0^2 + a_0^2  c_0^2  x_1^2, \quad L_1 = b_0^3  f_0 + a_0^3  e_0  x_1^2 \\
    & L_2 = 2  a_0  b_0  c_0  d_0  x_1, \quad L_3 = \sqrt{L_2 + L_1} \\
    & L_4 = \sqrt{L_1 - L_2}, \quad L_5 = a_0^2  d_0^2 + b_0^2  c_0^2  x_1^2 \\
    & L_6 = \sqrt{L_0 + L_2}, \quad L_7 = \sqrt{L_0 - L_2} \\
    & L_8 = \tan\left(\frac{\delta_1^{-}}{2}\right), \, \, L_9 = \tan\left(\frac{\delta_1^{+}}{2}\right), \, \, L_{10}= \tan\left(\frac{\delta_1^+ - \delta_1^-}{4}\right)\\
    & T_1 = \arctan\left(\frac{L_3 L_9}{L_4}\right) +  N^{+} \pi - \arctan\left(\frac{L_3 L_8}{L_4}\right) - N^{-} \pi \\
    & T_2 = \arctan\left(\frac{L_6 L_9}{L_7}\right) + N^{+} \pi - \arctan\left(\frac{L_6 L_8}{L_7}\right) - N^{-} \pi \\
    & R^\text{RP01}_{\Delta d} = 1 - \frac{2}{(\delta_1^+ - \delta_1^-) \cdot (L_1 - L_0)} \left[ \frac{g_0 x_1^2}{L_3 L_4} T_1 + \right. \label{eq:R_RP01_swan_delta} \\
    &\quad \quad \left. \frac{(L_0 - L_5) (L_1 - L_0) - g_0 x_1^2}{L_6 L_7} T_2 \right] \notag \\
    & T_\text{1M} = \arctan \left( \frac{L_4 L_{10}}{L_3} \right) + N_{\delta} \pi, \, \, T_\text{2M} = \arctan \left( \frac{L_7 L_{10}}{L_6} \right) + N_{\delta} \pi \notag \\
    & T_\text{1m} = \arctan \left( \frac{L_3 L_{10}}{L_4} \right) + N_{\delta} \pi, \, \, T_\text{2m} = \arctan \left( \frac{L_6 L_{10}}{L_7} \right) + N_{\delta} \pi \notag \\
    & R^\text{RP01}_{\Delta d, \, \text{M/m}} = 1 - \frac{4}{(\delta_1^+ - \delta_1^-) (L_1 - L_0)} \left( \frac{g_0 x_1^2}{L_3 L_4} T_\text{1M/1m} \right. \notag \\
    & \quad \quad + \left. \frac{(L_0 - L_5)(L_1 - L_0) - g_0 x_1^2}{L_6 L_7} T_\text{2M/2m} \right)
\end{align}
}
\label{eq:swanRdelta_total}
\end{subequations}
It should be noted that we have now added the corresponding correction factors to the original reflection formula and its envelopes, accounting also for the strongly-wedge shaped films. Figs. \ref{fig:swanepoel}d display the reflectance for a highly-tilted film. In this spectral range, one can also see one shift of the lower and upper envelopes for the reflectance spectrum.

\section{Substrate transmission and reflection}
\label{sec:fitting}

The thin film is typically deposited on a commercial glass substrate, which is relatively thick (with a known $d_2 \sim 1$ mm) and has plane-parallel surfaces, ensuring uniformity. Before performing the optical characterization of the film under study, it is a common practice to first characterize the optical properties of the substrate $(n_2, k_2)$ using transmittance and/or reflectance measurements of the substrate alone, denoted $\{T_\text{s}^\text{exp}(\lambda_i), R_\text{s}^\text{exp}(\lambda_i)\}$. This preliminary step allows us to counteract the influence of the substrate on the overall transmittance and reflectance of the sample, allowing a precise determination of the desired properties of the thin film, namely $(n_1, k_1, d_1)$.

Under the approximation $k_2 = 0$, the thickness of the substrate $d_\text{s}$ becomes irrelevant for the calculation of the transmission and reflection intensities, as the glass substrate will not absorb light. However, the real refractive index $n_2$ remains important because it determines the amount of light reflected on the film-to-substrate (interface 02) and substrate-to-air (interface 20), as seen in Fig. \ref{fig:model}, and described in Section 4.A. 

\begin{figure}[b!]
  \centering
   \includegraphics[width=\linewidth]{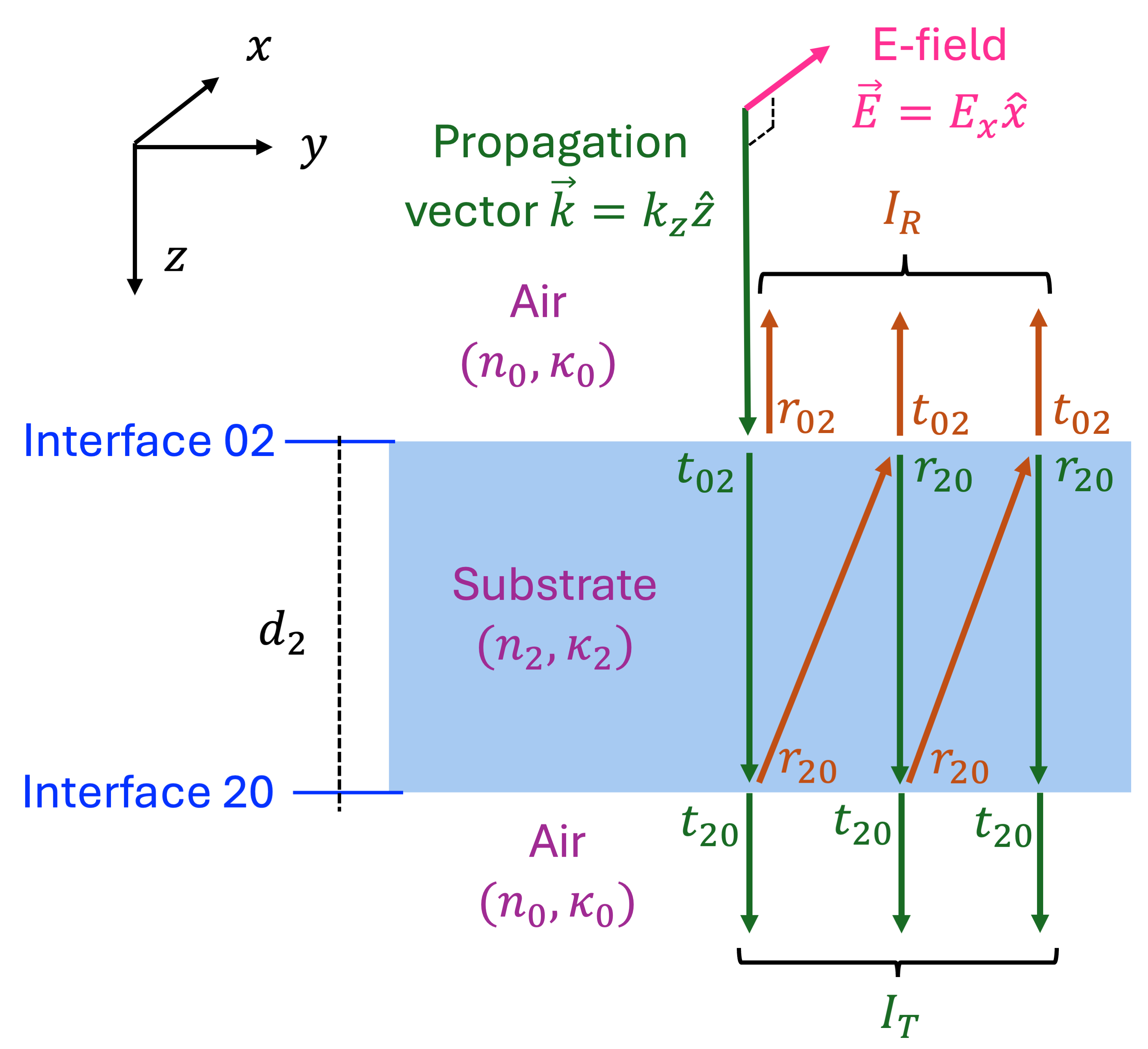}
   \caption{Diagram of the spectroscopic measurements for the substrate alone. The measures of $T_\text{s}$ and $R_\text{s}$ allow us to extract the substrate properties $n_2$ and $k_2$.}
   \label{fig:slab}
\end{figure}

Figure \ref{fig:slab} shows a representative diagram of the spectroscopic measurements for the substrate alone. In the simplistic scenario in which $\kappa_2 = 0$, $n_2$ can be uniquely determined \cite{yeh1990optical} from either transmission-only or reflection-only measurements by using the following relations:
\begin{subequations}
{\small
\begin{align}
    & T_\text{s}^\text{approx} = \frac{2 n_2}{n_2^2 + 1}, \, \, n_2 = \frac{1}{T_\text{s}^\text{approx}} + \left( \frac{1}{(T_\text{s}^\text{approx})^2} - 1 \right)^{1/2} \label{eq:Ts_approx} \\
    & R_\text{s}^\text{approx} = \frac{(n_2 - 1)^2}{n_2^2 + 1}, \, \, n_2 = \frac{1 + \sqrt{R_\text{s}^\text{approx} (2 - R_\text{s}^\text{approx})}}{1 - R_\text{s}^\text{approx}} \label{eq:Rs_approx} 
\end{align}
}
\end{subequations}

When the glass substrate \textit{does} absorb light, both transmission and reflection measurements of the substrate are necessary to uniquely determine $n_2(\lambda)$, $k_2(\lambda)$, and $d_\text{s}$. The formulae can be found using the same Abele transfer matrix formalism as in Section 4, and then using the \textit{spectral averaging} method to account for the limited coherence length, $L < 2 d_\text{s}$. However, for the particular case of a simple slab surrounding by air, it becomes more convenient to use the traditional technique of infinite incoherent summation \cite{nichelatti2002complex}: 
\begin{subequations}
{\small
\begin{align}
    & T_\text{s} = T_{02}^2 e^{-\alpha_2 d_2} \times \displaystyle \sum_{m=0}^{\infty} \big[ R_{02} e^{- \alpha_2 d_2} \big]^{2 m} = \label{eq:Ts} \\
    & \quad \quad \frac{T_{02}^2 e^{- \alpha_2 d_2}}{1 - R_{02}^2 e^{-2 \alpha_2 d_2}} \notag \\
    & R_\text{s} = R_{02} + T_{02}^2 R_{02} e^{-2 \alpha_2 d_2} \times \displaystyle \sum_{m=0}^{\infty} \big[ R_{02} e^{- \alpha_2 d_2} \big]^{2 m} = \label{eq:Rs}\\
    & \quad \quad R_{02} + \frac{R_{02} \, T_{02}^2 e^{-2 \alpha_2 d_2}}{1 - R_{02}^2 e^{-2 \alpha_2 d_2}} \notag 
\end{align}
}
\end{subequations}
Here, $R_{02} = |\mathbf{r}_{02}|^2 = |\mathbf{r}_{20}|^2$  and $T_{02} = |\mathbf{t}_{02}|^2 = |\mathbf{t}_{20}|^2$ represent the square norm of the Fresnel's coefficients\footnote{From Eqs. \ref{eq:fresnel}, we can see that, at normal incidence, the normed square of these coefficients is the same for the air-to-substrate and substrate-to-air interfaces}. Following the diagram from Fig. \ref{fig:slab}, we can see that the first addend ``$R_{02}$'' from Eq. \ref{eq:Rs} corresponds to the first bounce back of the light beam at the air-to-substrate interface. The second addend (for $m=0$) occurs when: \textit{(i)} the light passes through the air-to-substrate interface ($T_{02}$), \textit{(ii)} it then gets reflected at the substrate-to-air interface ($R_{02}$), and \textit{(iii)} it is finally transmitted through the substrate-to-air  ($T_{20}=T_{02}$) interface bouncing back to the detector above. During the round-trip of the beam within the slab, the Lambert-Beer law describes the amount of light absorption; this absorption is the responsible for the exponential term for $m = 0$ in the Eq. \ref{eq:Rs}. Further interreflections (for $m\ge 1$) just include other addends that can be reasoned in a similar fashion.

These \textit{direct} formulae that define $T_\text{s}$ and $R_\text{s}$ as functions of the substrate optical properties have been known for more than a century. However, determining the substrate properties $(n_2, k_2, d_2)$ from $T_\text{s}$ and $R_\text{s}$ is a more intricate \textit{inverse} problem. The work by Nichelatti \cite{nichelatti2002complex} in 2002 provided the first-ever \textit{analytical} expressions for that scenario. The following relations were found:
\begin{subequations}
{\small
\begin{align}
    & R_{02}(T_\text{s}, R_\text{s}) = \frac{2 + T_\text{s}^2 - (1 - R_\text{s})^2}{2 (2 - R_\text{s})} - \\ 
    & \quad \quad \frac{\sqrt{\big( 2 + T_\text{s}^2 - (1 - R_\text{s})^2 \big)^2 - 4 R_\text{s} (2 - R_\text{s})}}{2 (2 - R_\text{s})} \notag \\
    & k_2 = \frac{\lambda}{4 \pi h} \ln \left( \frac{R_{02} T_\text{s}}{R_\text{s} - R_{02}} \right) \label{eq:k2_from_TR} \\
    & n_2 = \frac{1 + R_{02}}{1 - R_{02}} \pm \sqrt{\frac{4 R_{02}}{(1 - R_{02})^2} - k_2^2} \label{eq:n2_from_TR}
\end{align}
}
\end{subequations}
Both the positive and negative signs in Eq. \ref{eq:n2_from_TR} define correct mathematical solutions to Eq. \ref{eq:Rs}. However, only one solution makes physical sense. While it is possible to discriminate the correct solution on a case-by-case basis by using commonly reported values, in our particular spectroscopic analysis, only the \textit{positive} solution is physically meaningful. Note that the film can be analyzed in spectral regions of high absorption (when 
$\kappa_1^2 \not < n_1^2$), as it can be very thin and still permit some transmission. However, using fully absorbing substrates with a high $\kappa_2$ in the spectral region of interest is \textit{not} desirable, as their significant thickness would then prevent any transmission entirely. Therefore, $\kappa_2^2 \ll n_2^2$ is always the practical scenario for the case of the substrate. That implies that $n_2$ from Eq. \ref{eq:n2_from_TR} is greater than one for the positive square-root (the correct physical solution) and less than one for the negative square-root (which must be discarded).

We can now compare the approximate formulae (with $\kappa_2 = 0$) from Eqs. \ref{eq:Ts_approx} and \ref{eq:Rs_approx} with the exact expressions from Eqs. \ref{eq:Ts} and \ref{eq:Rs}. It is important to note that for an absorbing substrate with $\kappa_2 = 10^{-6}$, the RMSE of the approximated formulae is 0.3\% for transmittance and 0.2\% for reflectance. Deriving $n_2$ from the simpler Eqs. \ref{eq:Ts_approx} and \ref{eq:Rs_approx} leads to an RMSE of 0.010\% and 0.005\%, respectively, for our simulated data.

\section{Concluding remarks}
In the present article, we derived the formulae for transmittance and reflectance of a normal-incident quasi-coherent light beam in a sample with a thin film, possibly with a high wedge profile, deposited on a thick absorbing substrate. Our model assumes a homogeneous, isotropic, and non-magnetic film material with linear response. 

In addition, we review other relevant formulae commonly used in the literature, examining their approximations and the level of accuracy. Rougher approximations lead to formulae that are easier to implement and compute. Then, a trade-off can be found between the accuracy of more complex expressions and the efficiency of approximate ones. Depending on the specific sample and the desired accuracy level, one can choose the most suitable model. For instance, for many conveniently-prepared films with uniform-thickness analyzed in the medium-to-low-absorption region, the simple Swanepoel formulae provide good accuracy with minimal computation time. However, our newly-developed formula proves to be more accurate in other more complicated scenarios: For instance, it is particularly effective when dealing with an `exotic' film surface, modeled with a large wedge parameter, $\Delta d > \lambda/(4 n)$, over a small illumination spot. Additionally, it is beneficial for the analysis across wide spectral ranges that encompass regions with strong film absorption ($\kappa_1^2 \not < n_1^2$) and substrate absorption ($\kappa_2 >0$).

A review of the formulae analyzed in our work can be found in Table \ref{tab:summary}. These expressions have been written with a consistent notation, tested numerically and analytically, and compared with each other. These formulae have been coded both in Python and in Matlab, and they are publicly available \href{https://drive.google.com/drive/folders/1Mv0p9or5ePowgt37yitNnw2Xe449IFTG?usp=sharing}{link}.

\newpage

\textbf{Appendix} 

Following the notation from Swanepoel \cite{swanepoel1989transmission}, we define the following coefficients:
\begin{subequations}
{\small 
\begin{align}
    \mathbf{r}_\text{ij} &= r_\text{ij} + i r_\text{ij}' \notag \\
    r_\text{ij} &= \frac{(n_\text{i}^2 - n_\text{j}^2) + (\kappa_\text{i}^2 - \kappa_\text{j}^2)}{(n_\text{i} + n_\text{j})^2 + (\kappa_\text{i} + \kappa_\text{j})^2} \notag \\
    r_\text{ij}' &= \frac{2(n_\text{i} \kappa_\text{j} - n_\text{j} \kappa_\text{i})}{(n_\text{i} + n_\text{j})^2 + (\kappa_\text{i} + \kappa_\text{j})^2} \notag \\
    R_\text{ij} &= |\mathbf{r}_\text{ij}|^2 = r_\text{ij}^2 + r_\text{ij}'^2 \notag \\
        a &= 1 + R_{01} R_{12} x_1^2 + R_{01} R_{23} x_1^2 x_2^2 + R_{12} R_{23} x_2^2 + \notag\\
         &\quad 2(r_{01} r_{12} (1 + R_{23} x_2^2) - r_{01}' r_{12}' (1 - R_{23} x_2^2)) x_1 \cos(\phi_1) + \notag \\
         &\quad 2(r_{01} r_{12}' (1 - R_{23} x_2^2) + r_{01}' r_{12} (1 + R_{23} x_2^2)) x_1 \sin(\phi_1) \notag \\
    e &= R_{01} + R_{12} x_1^2 + R_{23} x_1^2 x_2^2 + R_{01} R_{12} R_{23} x_2^2 + \notag \\
         &\quad 2(r_{01} r_{12} (1 + R_{23} x_2^2) + r_{01}' r_{12}' (1 - R_{23} x_2^2)) x_1 \cos(\phi_1) + \notag \\
         &\quad 2(r_{01} r_{12}' (1 - R_{23} x_2^2) - r_{01}' r_{12} (1 + R_{23} x_2^2)) x_1 \sin(\phi_1) \notag \\
    b &= r_{12} r_{23} (1 + R_{01} x_1^2) - r_{12}' r_{23}' (1 - R_{01} x_1^2) x_2 + \notag \\
         &\quad (r_{01} r_{23} (1 + R_{12}) - r_{01}' r_{23}' (1 - R_{12})) x_1 x_2 \cos(\phi_1) + \notag \\
         &\quad (r_{01} r_{23}' (1 - R_{12}) + r_{01}' r_{23} (1 + R_{12})) x_1 x_2 \sin(\phi_1) \notag \\
    f &= r_{12} r_{23} (x_1^2 + R_{01}) x_2 + r_{12}' r_{23}' (x_1^2 - R_{01}) x_2 + \notag \\
         &\quad (r_{01} r_{23} (1 + R_{12}) + r_{01}' r_{23}' (1 - R_{12})) x_1 x_2 \cos(\phi_1) + \notag \\
         &\quad (r_{01} r_{23}' (1 - R_{12}) - r_{01}' r_{23} (1 + R_{12})) x_1 x_2 \sin(\phi_1) \notag \\
    c &= (r_{12} r_{23}' (1 + R_{01} x_1^2) + r_{12}' r_{23} (1 - R_{01} x_1^2)) x_2 + \notag \\
         &\quad (r_{01} r_{23}' (1 + R_{12}) + r_{01}' r_{23} (1 - R_{12})) x_1 x_2 \cos(\phi_1) - \notag \\
         &\quad (r_{01} r_{23} (1 - R_{12}) - r_{01}' r_{23}' (1 + R_{12})) x_1 x_2 \sin(\phi_1) \notag \\
    g &= r_{12} r_{23}' (x_1^2 + R_{01}) x_2 - r_{12}' r_{23} (x_1^2 - R_{01}) x_2 + \notag \\
         &\quad (r_{01} r_{23}' (1 + R_{12}) - r_{01}' r_{23} (1 - R_{12})) x_1 x_2 \cos(\phi_1) - \notag \\
         &\quad (r_{01} r_{23} (1 - R_{12}) + r_{01}' r_{23}' (1 + R_{12})) x_1 x_2 \sin(\phi_1) \notag \\
    u &= 1 + R_{01} R_{12} x_1^2 - R_{01} R_{23} x_1^2 x_2^2 - R_{12} R_{23} x_2^2 + \notag \\
         &\quad 2(r_{01} r_{12} (1 - R_{23} x_2^2) - r_{01}' r_{12}' (1 + R_{23} x_2^2)) x_1 \cos(\phi_1) + \notag \\
         &\quad 2(r_{01} r_{12}' (1 + R_{23} x_2^2) + r_{01}' r_{12} (1 - R_{23} x_2^2)) x_1 \sin(\phi_1) \notag \\
    w &= 1 + R_{01} R_{12} x_1^2 + 2(r_{01} r_{12} - r_{01}' r_{12}') x_1 \cos(\phi_1) + \notag \\
         &\quad 2(r_{01} r_{12}' + r_{01}' r_{12}) x_1 \sin(\phi_1) \notag \\
    h &= ((1 + r_{01})^2 + r_{01}'^2) ((1 + r_{12})^2 + r_{12}'^2) ((1 + r_{23})^2 + r_{23}'^2) x_1 x_2 \notag \\
a'_{1} &= 1 + R_{01} R_{12} x_{1}^{2} - R_{01} R_{23} x_{1}^{2} x_{2}^{2} - R_{12} R_{23} x_{2}^{2} \notag \\
b'_{1} &= \left( r_{01} r_{12} (1 - R_{23} x_{2}^{2}) - r'_{01} r'_{12} (1 + R_{23} x_{2}^{2}) \right) x_{1} \notag \\
c'_{1} &= \left( r_{01} r'_{12} (1 + R_{23} x_{2}^{2}) + r'_{01} r_{12} (1 - R_{23} x_{2}^{2}) \right) x_{1} \notag
\end{align}
}
\label{eq:fresnel}
\end{subequations}

\vspace{0.5 cm}

\newpage

\section*{Bibliography}
\bibliography{refs}

\end{document}